\documentclass[12pt]{article}
\usepackage{color}
\usepackage{graphicx}
\usepackage{amsmath}
\usepackage{times}
\usepackage{subfigure}
\usepackage{cite}
\usepackage{amssymb}
\usepackage{amsthm}

\newtheorem{algorithm}{Algorithm}
\def\diag{{\rm diag}}
\def\bx{{\bf x}}
\def\by{{\bf y}}
\def\bz{{\bf z}}
\def\bc{{\bf c}}
\def\bb{{\bf b}}
\def\ba{{\bf a}}

\def\Exp{{\rm E}}
\def\diag{{\rm diag}}
\def\Prob{{\rm P}}

\newtheorem{theorem}{Theorem}
\newtheorem{lemma}{Lemma}

\title{Network Beamforming Using Relays with Perfect Channel Information}
\author{
\textsc{Yindi Jing and Hamid Jafarkhani}\thanks{This work was supported in part by ARO under the Multi-University
Research Initiative (MURI) grant \#W911NF-04-1-0224.}
\\
{University of California, Irvine, Irvine, CA, 92697 } \\
}
\date{      }
\begin{document}
\maketitle

\begin{abstract}
This paper is on beamforming in wireless relay networks with perfect channel information at relays, the receiver, and the transmitter if there is a direct link between the transmitter and receiver. It is assumed that every node in the network has its own power constraint. A two-step amplify-and-forward protocol is used, in which the transmitter and relays not only use match filters to form a beam at the receiver but also adaptively adjust their transmit powers according to the channel strength information. For a network with any number of relays and no direct link, the optimal power control is solved analytically. The complexity of finding the exact  solution is linear in the number of relays. Our results show that the transmitter should always use its maximal power and the optimal power used at a relay is not a binary function. It can take any value between zero and its maximum transmit power. Also, surprisingly, this value depends on the quality of all other channels in addition to the relay's own channels. Despite this coupling fact, distributive strategies are proposed in which, with the aid of a low-rate broadcast from the receiver, a relay needs only its own channel information to implement the optimal power control. Simulated performance shows that network beamforming achieves the maximal diversity and outperforms other existing schemes.

Then, beamforming in networks with a direct link are considered. We show that when the direct link exists during the first step only, the optimal power control at the transmitter and relays is the same as that of networks with no direct link. For networks with a direct link during the second step only and both steps, recursive numerical algorithms are proposed to solve the power control problem. Simulation shows that by adjusting the transmitter and relays' powers adaptively, network performance is significantly improved.
\end{abstract}

\section{Introduction}
It is well-known that due to the fading effect, the transmission over wireless channels suffers from severe attenuation in signal strength. Performance of wireless communication is much worse than that of wired communication. For the simplest point-to-point communication system, which is composed of one transmitter and one receiver only, the use of multiple antennas can improve the capacity and reliability. Space-time coding and beamforming are among the most successful techniques developed for multiple-antenna systems during the last decades \cite{Jbook,HTW}. However, in many situations, due to the limited size and processing power, it is not practical for some users, especially small wireless mobile devices, to implement multiple antennas. Thus, recently, wireless network communication is attracting more and more attention. A large amount of effort has been given to improve the communication by having different users in a network cooperate. This improvement is conventionally addressed as cooperative diversity and the techniques cooperative schemes.

Many cooperative schemes have been proposed in literature \cite{HuaMeiChangAsil,HuaMeiChang,TangVal,SenErkipAa1,SenErkipAa2,NaBoKn,BoNaOyPa,DSTC-paper,DanaHassibi,LenemanWornell,JHHN,DDSTC,kira2,OgHa-allerton,zz-eg-sc,LaTsWo,KaSh,HuSaNo,DeMiTa}. Some assume channel information at the receiver but not the transmitter and relays, for example, the noncoherent amplify-and-forward protocol in \cite{NaBoKn,BoNaOyPa} and distributed space-time coding in \cite{DSTC-paper}. Some assume channel information at the receiving side of each transmission, for example, the decode-and-forward protocol in \cite{LenemanWornell,NaBoKn} and the coded-cooperation in \cite{JHHN}. Some assume no channel information at any node, for example, the differential transmission schemes proposed independently in \cite{DDSTC,kira2,OgHa-allerton}. The coherent amplify-and-forward scheme in \cite{DanaHassibi,BoNaOyPa} assumes full channel information at both relays and the receiver. But only channel direction information is used at relays. In all these cooperative schemes, the relays always cooperate on their highest powers. None of the above pioneer work allow relays to adjust their transmit powers adaptively according to channel magnitude information, and this is exactly the concern of this paper. 

There have been several papers on relay networks with adaptive power control. In \cite{LaCa,CVS}, outage capacity of networks with a single relay and perfect channel information at all nodes were analyzed. Both work assume a total power constraint on the relay and the transmitter. A decode-and-forward protocol is used at the relay, which results in a binary power allocation between the relay and the transmitter. In \cite{lar}, performance of networks with multiple amplify-and-forward relays and an aggregate power constraint was analyzed. A distributive scheme for the optimal power allocation is proposed, in which each relay only needs to know its own channels and a real number that can be broadcasted by the receiver. Another related work on networks with one and two amplify-and-forward relays can be found in \cite{AnKa}. In \cite{AKSA-Comm-06}, outage minimization of single-relay networks with limited channel-information feedback is performed. It is assumed that there is a long-term power constraint on the total power of the transmitter and the relay. In this paper, we consider networks with a general number of amplify-and-forward relays and we assume a separate power constraint on each relay and the transmitter. Due to the difference in the power assumptions, compared to \cite{lar}, analysis of this new model is more difficult and totally different results are obtained.

For multiple-antenna systems, when there is no channel information at the transmitter, space-time coding can achieve full diversity \cite{Jbook}. If the transmitter has perfect or partial channel information, performance can be further improved through beamforming since it takes advantage of the channel information (both direction and strength) at the transmit side to obtain higher receive SNR \cite{HTW}. With perfect channel information or high quality channel information feedback from the receiver at the transmitter, one-dimensional beamforming is proved optimal \cite{HTW,NLTW,JaGo}. The more practical multiple-antenna systems with partial channel information at the transmitter, channel statistics or quantized instantaneous channel information, are also analyzed extensively \cite{msea,rora,rora06,ZhGi-mean,ZhGi-correlation}. In many situations, appropriate combination of beamforming and space-time coding outperforms either one of the two schemes alone \cite{jso,Liu-Jar-SP,LoHe,EkJa}. In this paper, we will see similar performance improvement in networks using network beamforming over distributed space-time coding and other existing schemes such as best-relay selection and coherent amplify-and-forward.

We consider networks with one pair of transmitter and receiver but multiple relays. The receiver knows all channels and every relay knows its own channels perfectly. In networks with a direct link (DL) between the transmitter and the receiver, we also assume that the transmitter knows the DL fully. A two-step amplify-and-forward protocol is used, where in the first step, the transmitter sends information and in the second step, the transmitter and relays, if there is a DL, transmit. We first solve the power control problem for networks with no DL analytically. The exact solution can be obtained with a complexity that is linear in the number of relays. Then, to perform network beamforming, we propose two distributive strategies in which a relay needs only its own channel information and a low-rate broadcast from the receiver. Simulation shows that the optimal power control or network beamforming outperforms other existing schemes. We then consider networks with a DL during the first transmission step, the second transmission step, and both. For the first case, the power control problem is proved to be the same as the one in networks without the DL. For the other two cases, recursive numerical algorithms are provided. Simulation shows that they have much better performance compared to networks without power control. We should clarify that only amplify-and-forward is considered here. For decode-and-forward, the result may be different and it depends on the details of the coding schemes. 

The paper is organized as follows. In the next section, the relay network model and the main problem are introduced. Section \ref{sec-PL} works on the power control problem in relay networks with no DL and Section \ref{sec-DL} considers networks with a DL. Section \ref{sec-conclusion} contains the conclusion and several future directions.

\section{Wireless Relay Network Model and Problem Statement}

\begin{figure}[t]
\centering
\includegraphics[width=4in]{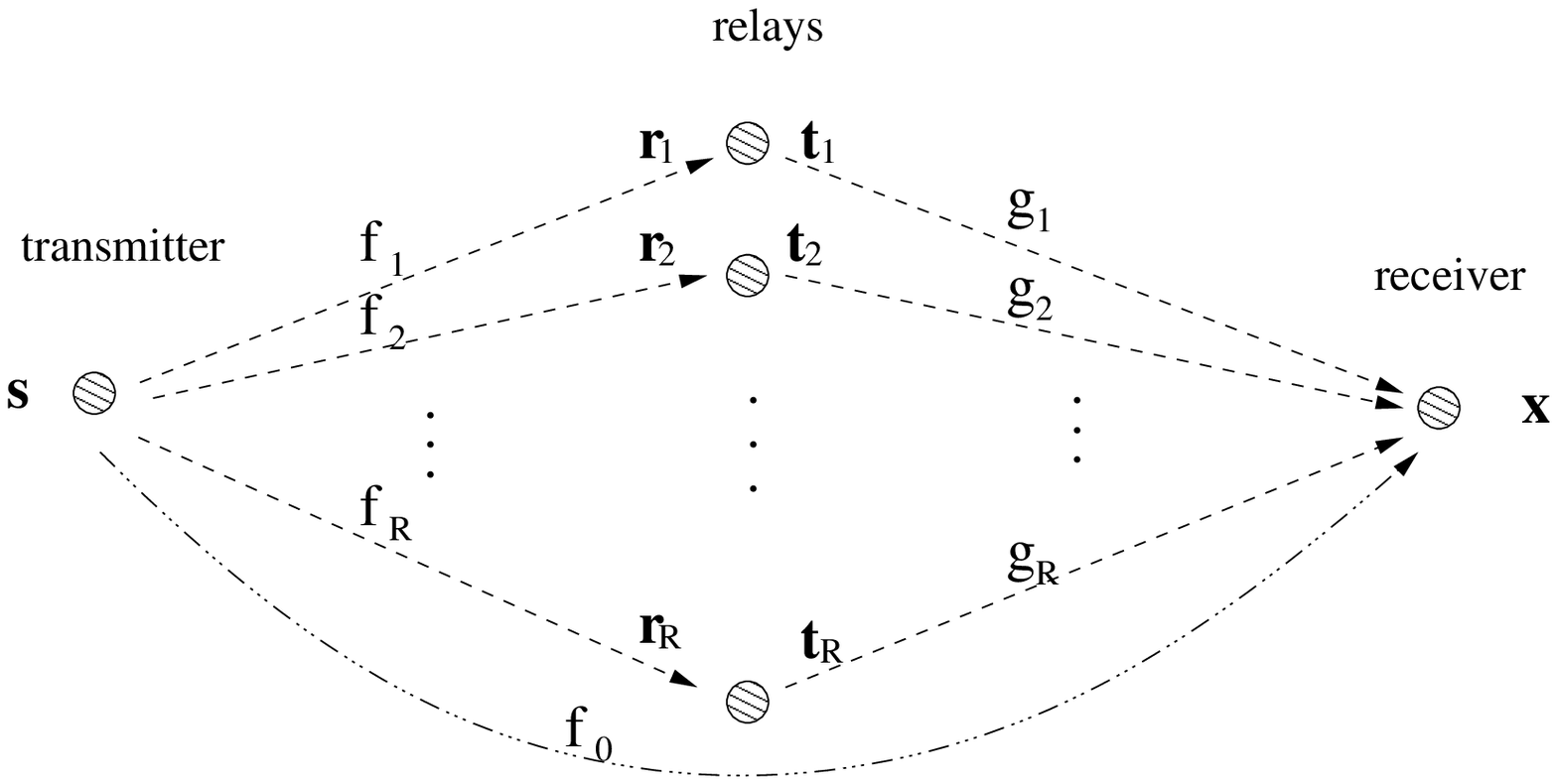}
\caption{Wireless relay network.}
\label{fig-network}
\end{figure}

Consider a relay network with one transmit-and-receive pair and $R$ relays as depicted in Fig.~\ref{fig-network}. Every relay has only one single antenna which can be used for both transmission and reception. Denote the channel from the transmitter to the $i$th relay as $f_i$ and the channel from the $i$th relay to the receiver as $g_i$. If the DL between the transmitter and the receiver exists, we denote it as $f_0$. We assume that the transmitter knows $f_0$, the $i$th relay knows its own channels $f_i$ and $g_i$, and the receiver knows all channels $f_0,f_1,\dots,f_R$ and $g_1,\dots,g_R$. The channels can have both fading and path-loss effects. Actually, our results are valid for any channel statistics. We assume that for each transmission, the powers used at the transmitter and the $i$th relay are no larger than $P_0$ and $P_i$, respectively. Note that in this paper, only short-term power constraint is considered, that is, there is an upper bound on the average transmit power of each node for each transmission. A node cannot save its power to favor transmissions with better channel realizations.

We use a two-step amplify-and-forward protocol. During the first step, the transmitter sends $\alpha_0\sqrt{P_0}s$. The information symbol $s$ is selected randomly from the codebook $\mathcal{S}$. If we normalize it as $\Exp |s|^2=1 $, the average power used at the transmitter is $\alpha_0^2P_0$. The $i$th relay and the receiver, if a DL exists during the first step, receive 
\begin{equation}
r_i=\alpha_0\sqrt{P_0}f_is+v_i \hspace{2mm} \mbox{and} \hspace{2mm} x_1=\alpha_0\sqrt{P_0}f_0s+w_1,
\label{x1}
\end{equation}
respectively. $v_i$ and $w_1$ are the noises at the $i$th relay and the receiver at Step 1. We assume that they are $\mathcal{CN}(0,1)$. During the second step, the transmitter sends $\beta_0\sqrt{P_0}e^{j\theta_0}s$, if a DL exists during this step. At the same time, the $i$th relay sends 
\[
t_i=\alpha_i\sqrt{\frac{P_i}{1+\alpha_0^2|f_i|^2P_0}}e^{j\theta_i}r_i.
\]
The average transmit power of the $i$th relay can be calculated to be $\alpha_i^2P_i$. If we assume that $f_0$ keeps constant for the two steps, the receiver gets 
\setlength{\arraycolsep}{0.0em}
\begin{eqnarray}
x_2&=&\beta_0\sqrt{P_0}f_0e^{j\theta_0}s+\sum_{i=1}^Rg_it_i+w_2 \nonumber\\
&=&\sqrt{P_0}\left(\beta_0f_0e^{j\theta_0}+\alpha_0\sum_{i=1}^R \frac{\alpha_if_ig_ie^{j\theta_i}\sqrt{P_i}}{\sqrt{1+\alpha_0^2|f_i|^2P_0}}\right)s 
+\sum_{i=1}^R\frac{\alpha_ig_ie^{j\theta_i}\sqrt{P_i}}{\sqrt{1+\alpha_0^2|f_i|^2P_0}}v_i+w_2.
\label{x2-old}
\end{eqnarray}
\setlength{\arraycolsep}{0.5em}
$w_2$ is the noise at the receiver at Step 2, which is also assumed to be $\mathcal{CN}(0,1)$. Note that if the transmitter sends during both steps, we assume that the total average power it uses is no larger than $P_0$. With this, the total average power in transmitting one symbol is no larger than $\sum_{i=0}^RP_i$. Clearly, the coefficients $\alpha_0,\alpha_1,\dots,\alpha_R$ are introduced in the model for power control.  The power constraints at the transmitter and relays require that $\alpha_0^2+\beta_0^2\le 1$ and $0\le \alpha_i\le 1$.

Our network beamforming design is thus the design of $\theta_0,\theta_1,\cdots,\theta_R$ and $\alpha_0,\beta_0,\alpha_1,\cdots,\alpha_R$, such that the error rate of the network is the smallest. This is equivalent to maximize the receive SNR, or the total receive SNR of both branches if a DL exists during the first step. From (\ref{x2-old}), we can easily prove that an optimal choice of the angles are $\theta_0=-\arg f_0$ and $\theta_i=-(\arg f_i+\arg g_i)$. That is, match filters should be used at relays and the transmitter during the second step to cancel the phases of their channels and form a beam at the receiver. We thus have
\begin{equation}
x_2=\sqrt{P_0}\left(\beta_0|f_0|+\alpha_0\sum_{i=1}^R \frac{\alpha_i|f_ig_i|\sqrt{P_i}}{\sqrt{1+\alpha_0^2|f_i|^2P_0}}\right)s
+\sum_{i=1}^R\frac{\alpha_i|g_i|\sqrt{P_i}}{\sqrt{1+\alpha_0^2|f_i|^2P_0}}e^{-j\arg f_i}v_i+w_2.
\label{x2}
\end{equation}
What is left is the optimal power control, i.e., the choice of $\alpha_0,\beta_0,\alpha_1,\dots,\alpha_R$. This is also the main contribution of our work.

\section{Optimal Relay Power Control}
\label{sec-PL}
In this section, we investigate the optimal adaptive power control at the transmitter and relays in networks without a DL. Section \ref{sec-PL-result} presents the analytical power control result. Section \ref{sec-PL-discussion} comments on the result and gives distributive schemes for the optimal power control. Section \ref{sec-PL-simulation} provides simulated performance.

\subsection{Analytical Result}
\label{sec-PL-result}
With no DL, we have $\beta_0=0$ and $x_1=0$. From (\ref{x2}), the receive SNR can be calculated to be
\[
\frac{\alpha_0P_0\left(\sum_{i=1}^R\frac{\alpha_i|f_ig_i|\sqrt{P_i}}{\sqrt{1+\alpha_0^2|f_i|^2P_0}}\right)^2}
{1+\sum_{i=1}^R\frac{\alpha_i^2|g_i|^2P_i}{1+\alpha_0^2|f_i|^2P_0}}.
\]
It is an increasing function of $\alpha_0$. Therefore, the transmitter should always use its maximal power, i.e., $\alpha_0^*=1$. The receive SNR is thus:
\[
\frac{P_0\left(\sum_{i=1}^R\frac{\alpha_i|f_ig_i|\sqrt{P_i}}{\sqrt{1+|f_i|^2P_0}}\right)^2} {1+\sum_{i=1}^R\frac{\alpha_i^2|g_i|^2P_i}{1+|f_i|^2P_0}}.
\]

Before going into details of the SNR optimization, we first introduce some notation to help the presentation. $\langle \cdot,\cdot \rangle$ indicates the inner product. $\|\cdot\|$ indicates the 2-norm. $\Prob$ indicates the probability. $a_i$ denotes the $i$th coordinate of vector $\ba$ and $\ba_{i_1,\dots,i_k}$ denotes the $k$-dimensional vector $\left[\begin{array}{ccc} a_{i_1} & \cdots & a_{i_k}\end{array}\right]^T$, where ${\cdot}^T$ represents the transpose. If $\ba,\bb$ are two $R$-dimensional vectors, $\ba\preceq \bb$ means $a_i \le b_i$ for all $i=1,\dots,R$.  $0_R$ is the $R$-dimensional vector with all zero entries. Denote the set $0_R \preceq \by  \preceq \ba$ or equivalently, $0\le y_i\le a_i$ for $i=1,\dots,R$, as $\Lambda$. For $1\le k\le R-1$, denote the set $0_{k} \preceq \by_{i_{1},\dots,i_k} \preceq \ba_{i_{1},\dots,i_k}$ as $\Lambda_{i_1,\dots,i_k}$, where $\{i_1,\dots,i_k\}$ is a $k$-subset of $\{1,\dots,R\}$. 

Define 
\[
\bx=\left[\begin{array}{c} \alpha_1 \\ \vdots \\ \alpha_R \end{array}\right],
\bb=\left[\begin{array}{c} \frac{|f_1g_1|\sqrt{P_1}}{\sqrt{1+|f_1|^2P_0}} \\ \vdots \\ 
\frac{|f_Rg_R|\sqrt{P_R}}{\sqrt{1+|f_R|^2P_0}} \end{array} \right], 
\ba=\left[\begin{array}{c} \frac{|g_1|\sqrt{P_1}}{\sqrt{1+|f_1|^2P_0}} \\ \vdots \\ \frac{|g_R|\sqrt{P_R}}{\sqrt{1+|f_R|^2P_0}}\end{array}\right], 
\hspace{3mm} \mbox{and} \hspace{3mm} A=\diag\{\ba\},
\]
where $\diag\{\ba\}$ indicates the diagonal matrix whose $i$th diagonal entry is $a_i$. With the transformation $\by=A\bx$, or equivalently, $\bx=A^{-1}\by$, we have 
\[
SNR=P_0\frac{\langle\bb, \bx \rangle^2}{1+\|A\bx\|^2}
%=P_0\frac{\by^T\bc \bc ^T\by }{1+\by^T\by}
=P_0\frac{\langle\bc,\by \rangle^2}{1+\|\by\|^2},
\]
where 
\[
\bc=A^{-T}\bb=\left[\begin{array}{ccc}
\frac{\sqrt{1+|f_1|^2P_0}}{|g_1|\sqrt{P_1}} & \cdots & 0 \\
\vdots & \ddots & \vdots \\ 0 & \cdots & \cdots \frac{\sqrt{1+|f_R|^2P_0}}{|g_R|\sqrt{P_R}} \end{array}\right]
\left[\begin{array}{ccc}\frac{|f_1g_1|\sqrt{P_1}}{\sqrt{1+|f_1|^2P_0}} \\ \vdots \\ \frac{|f_Rg_R|\sqrt{P_R}}{\sqrt{1+|f_R|^2P_0}}\end{array}\right]
=\left[\begin{array}{ccc}|f_1| \\ \vdots \\ |f_R| \end{array}\right].
\]
The receive SNR optimization problem is thus equivalent to
\begin{equation}
\max_{\by}\frac{\langle \bc, \by\rangle^2}{1+\|\by\|^2} \hspace{3mm} \mbox{s.t.} \hspace{3mm} \by\in\Lambda.
\label{opt-problem}
\end{equation}
The difficulty of the problem lies in the shape of the feasible set. If $\by$ is constrained on a hypersphere, that is, $\|\by\|=r$, the solution is obvious at least geometrically. Given that $\|\by\|=r$, 
\[
\frac{\langle \bc, \by\rangle^2}{1+\|\by\|^2}=\frac{r^2\|\bc\|^2}{1+r^2}\cos^2\varphi,
\]
where $\varphi$ is the angle between $\bc$ and $\by$. The optimal solution should be the vector which has the smallest angle with $\bc$. Thus, we decompose (\ref{opt-problem}) as
\begin{equation}
\max_{r}\frac{1}{1+r^2} \left(\max_{\|\by\|=r}{\langle \bc, \by\rangle}\right)^2 \hspace{3mm} \mbox{s.t.} \hspace{3mm} \by\in\Lambda \hspace{2mm}\mbox{and} \hspace{2mm} 0\le r \le \|\ba\|. 
\label{opt-problem-1}
\end{equation}
Since $\Prob(a_i>0)=1$ and $\Prob(c_i>0)=1$, we assume that $a_i>0$ and $c_i>0$. Define 
\begin{equation}
\phi_j=\phi(f_j,g_j,P_j)=\frac{c_j}{a_j}=\frac{|f_j|\sqrt{1+|f_j|^2P_0}}{|g_j|\sqrt{P_j}},
\label{phi}
\end{equation}
for $i=1,\dots,R$ and, for the sake of presentation, define $\phi_{R+1}=0$. Order $\phi_j$ as
\begin{equation}
\phi_{\tau_1}\ge \phi_{\tau_2}\ge \cdots \ge \phi_{\tau_R}\ge \phi_{\tau_{R+1}}.
\label{order}
\end{equation}
$(\tau_1,\tau_2,\dots,\tau_R,\tau_{R+1})$ is thus an ordering of $(1,2,\dots,R,R+1)$ and $\tau_{R+1}=R+1$. Define
\setlength{\arraycolsep}{0.0em}
\begin{eqnarray*}
r_0&=&0, \nonumber \\
r_1&=&\phi_{\tau_1}^{-1}\|\bc\|=\sqrt{\phi_{\tau_1}^{-2}\|\bc_{\tau_2,\dots,\tau_R}\|^2+a_{\tau_1}^2}, \nonumber\\
r_2&=&\sqrt{\phi_{\tau_2}^{-2}\|\bc_{\tau_2,\dots,\tau_R}\|^2+a_{\tau_1}^2}
=\sqrt{\phi_{\tau_2}^{-2}\|\bc_{\tau_3,\dots,\tau_R}\|^2+\sum_{i=1}^2a_{\tau_i}^2}, \nonumber \\
&\vdots& \nonumber \\
r_{R-1}&=&\sqrt{\phi_{\tau_{R-1}}^{-2}\|\bc_{\tau_{R-1},\tau_R}\|^2+\sum_{i=1}^{R-2}a_{\tau_i}^2}
=\sqrt{\phi_{\tau_{R-1}}^{-2}|c_{\tau_R}|^2+\sum_{i=1}^{R-1}a_{\tau_i}^2},\nonumber \\
r_R&=&\sqrt{\phi_{\tau_R}^{-2}|c_{\tau_R}|^2+\sum_{i=1}^{R-1}a_{\tau_i}^2}=\|\ba\|.
%\label{radius}
\end{eqnarray*}
\setlength{\arraycolsep}{0.5em}
Since $\phi_{\tau_{j-1}}\ge \phi_{\tau_{j}}$, we have $r_{j-1}\le r_j$ for $j=1,\dots,R$. Thus, the feasible interval of the radius, $[0,\|\ba\|]$, can be decomposed into the following $R$ intervals:
\[
[0,\|\ba\|]=[r_0,r_1]\cup [r_1,r_2]\cup \cdots\cup [r_{R-2},r_{R-1}]\cup [r_{R-1},r_R].
\]
We denote $\Gamma_i=[r_i,r_{i+1}]$ for $i=0,\dots,R-1$. Thus, (\ref{opt-problem-1}) is equivalent to
\[
\max_{i=1,\dots,R}\max_{r\in\Gamma_i}\frac{1}{1+r^2} \left(\max_{\|\by\|=r\in\Gamma_i,\by\in\Lambda}{\langle \bc, \by\rangle}\right)^2.
%\label{decom-problem}
\]
We have decomposed the optimization problem into $R$ subproblems. We now work on the $i$th subproblem:
\begin{equation}
\max_{r\in\Gamma_i}\frac{1}{1+r^2} \left(\max_{\|\by\|=r\in\Gamma_i,\by\in\Lambda}{\langle \bc, \by\rangle}\right)^2.
\label{sub-problem-i}
\end{equation}
Denote the solution of the inner optimization problem,
\begin{equation}
\max_{\|\by\|=r\in\Gamma_i,\by\in\Lambda} \langle \bc, \by\rangle,
\label{inner-problem}
\end{equation}
as $\bz^{(i)}$. We have the following two lemmas.

\begin{lemma}
$z^{(i)}_j=a_j$ for $j=\tau_1,\dots, \tau_i$. 
\label{lemma-edge}
\end{lemma}

\begin{proof}
We prove this lemma by contradiction. Assume that $z^{(i)}_j<a_j$ for some $j\in\{\tau_1,\dots, \tau_i\}$. We first show that there exists an $l\in \{\tau_{i+1},\dots, \tau_R\}$ such that $\frac{z^{(i)}_j}{c_j}<\frac{z^{(i)}_l}{c_l}$. Assume that $\frac{z^{(i)}_j}{c_j}\ge\frac{z^{(i)}_m}{c_m}$ for all $m\in \{\tau_{i+1},\dots, \tau_R\}$. We have $z^{(i)}_m \le c_m\frac{z^{(i)}_j}{c_j}<c_m\frac{a_j}{c_j}=c_m\phi_{j}^{-1}$. Thus, 
\begin{eqnarray*}
\|\bz^{(i)}\|&=&\sqrt{\sum_{m=1}^i \left(z^{(i)}_{\tau_m}\right)^2 +\sum_{m=i+1}^R \left(z^{(i)}_{\tau_m}\right)^2} \\
&<&\sqrt{\sum_{m=1}^i a_{\tau_m}^2 +\sum_{m=i+1}^R c_{\tau_m}^2 \phi_{j}^{-2}} \\
&=&\sqrt{\phi_{j}^{-2} \|c_{\tau_{i+1},\dots,\tau_R}\|^2+\sum_{m=1}^i a_{\tau_m}^2} \\
&\le& \sqrt{\phi_{\tau_i}^{-2} \|c_{\tau_{i+1},\dots,\tau_R}\|^2+\sum_{m=1}^{i} a_{\tau_m}^2} \hspace{10mm}
\mbox{because of (\ref{order})}\\
&=&r_i.
\end{eqnarray*}
This contradicts $\|\bz^{(i)}\|\in\Gamma_i$. thus, there exists an $l\in \{\tau_{i+1},\dots, \tau_R\}$ such that $\frac{z^{(i)}_j}{c_j}<\frac{z^{(i)}_l}{c_l}$. 

Define another vector $\bz'$ as $z'_j=z^{(i)}_j+\delta$, $z'_l=\sqrt{\left(z^{(i)}_l\right)^2-2\delta z^{(i)}_j-\delta^2}$, and $z'_m=z^{(i)}_m$ for $m\ne i,l$, where
\[
0<\delta<\min\left\{2c_j\left(1+\frac{c_j^2}{c_l^2} \right)^{-1}\left(\frac{z^{(i)}_l}{c_l} -\frac{z^{(i)}_j}{c_j}\right), \sqrt{\left(z^{(i)}_j\right)^2+\left(z^{(i)}_l\right)^2}-z^{(i)}_j,a_j-z^{(i)}_j\right\}.
\]
Since we have assumed that $z^{(i)}_j<a_j$ and have just proved that $\frac{z^{(i)}_j}{c_j}<\frac{z^{(i)}_l}{c_l}$, such $\delta$ is achievable. To contradict the assumption that $\bz^{(i)}$ is the optimal, it is enough to prove the following two items:
\begin{enumerate}
\item $\bz'$ is a feasible point: $\|\bz'\|=r$ and $\bz'\in\Lambda$,
\item $\langle \bc, \bz^{(i)}\rangle<\langle \bc, \bz'\rangle$.
\end{enumerate}
From the definition of $\bz'$, we have 
\[\|\bz'\|^2=\left(z'_j\right)^2+\left(z'_l\right)^2+\sum_{m\ne j,l}\left(z'_m\right)^2=\left(z^{(i)}_j+\delta\right)^2+\left(z^{(i)}_l\right)^2-2\delta z^{(i)}_j-\delta^2+\sum_{m\ne j,l}\left(z^{(i)}_m\right)^2=\|\bz^{(i)}\|^2=r^2.\]
Since $0<\delta<a_j-z^{(i)}_j$, we have $0<\bz'_j<a_j$. Also, since $0<\delta<\sqrt{\left(z^{(i)}_j\right)^2+\left(z^{(i)}_l\right)^2}-z^{(i)}_j$, we can easily prove that $z'_l=\sqrt{\left(z^{(i)}_l\right)^2-2\delta z^{(i)}_j-\delta^2}>0$ and $z'_l<z^{(i)}_l\le a_l$. Thus, $\bz'\in\Lambda$. The first item has been proved. For the second item, since $\delta<2c_j\left(1+\frac{c_j^2}{c_l^2} \right)^{-1}\left(\frac{z^{(i)}_l}{c_l} -\frac{z^{(i)}_j}{c_j}\right)$, we have
\begin{eqnarray*}
&&\left(1+\frac{c_j^2}{c_l^2} \right)\delta^2<2c_j\left(\frac{z^{(i)}_l}{c_l} -\frac{z^{(i)}_j}{c_j}\right)\delta 
= 2\left(z^{(i)}_l\frac{c_j}{c_l} -z^{(i)}_j\right)\delta\\
&\Rightarrow& \left(z^{(i)}_l\right)^2+\frac{c_j^2}{c_l^2}\delta^2-2\frac{c_j}{c_l}z^{(i)}_l\delta 
              <\left(z^{(i)}_l\right)^2-2z^{(i)}_j\delta-\delta^2 =z'^2_l\\
&\Rightarrow& c_lz^{(i)}_l-c_j\delta<c_l z'_l \\
%&\Rightarrow& c_lz^{(i)}_l-c_j (z'_j-z^{(i)}_j) - c_l z'_l<0 \\
&\Rightarrow&c_lz^{(i)}_l+c_jz^{(i)}_j-(c_l z'_l+c_j z'_j)<0 \\
&\Rightarrow& \langle \bc, \bz^{(i)}\rangle<\langle \bc, \bz'\rangle.
\end{eqnarray*}
%This contradicts our assumption that $\bz^{(i)}$ is optimal, thus finished our proof.
\end{proof}

\begin{lemma}
$z^{(i)}_j= \frac{\sqrt{r^2-\sum_{m=1}^ia_{\tau_m}^2}}{\|\bc_{\tau_{i+1},\dots,\tau_R}\|} c_j$ for $j=\tau_{i+1},\dots, \tau_R$.
\label{lemma-scale}
\end{lemma}

\begin{proof}
From Lemma \ref{lemma-edge}, $z^{(i)}_j=a_j$ for $j=\tau_1,\dots, \tau_i$. Thus, (\ref{inner-problem}) can be written as
\begin{eqnarray*}
&&\max_{\|\by\|=r\in\Gamma_i,\by\in\Lambda} \sum_{m=1}^ia_{\tau_m}c_{\tau_m}+ \langle \bc_{\tau_{i+1},\dots,\tau_R}, \by_{\tau_{i+1},\dots,\tau_R}\rangle \\
&=&\sum_{m=1}^ib_{\tau_m}+ \max_{{ \|\by_{\tau_{i+1},\dots,\tau_R}\|=\sqrt{r^2-\sum_{m=1}^ia^2_{\tau_m}},} \atop r\in\Gamma_i,\by_{\tau_{i+1},\dots,\tau_R}\in\Lambda_{\tau_{i+1},\dots,\tau_R}}  \langle \bc_{\tau_{i+1},\dots,\tau_R}, \by_{\tau_{i+1},\dots,\tau_R}\rangle.
\end{eqnarray*}
Define $\lambda=\frac{\sqrt{r^2-\sum_{m=1}^ia^2_{\tau_m}}}{\|\bc_{\tau_{i+1},\dots,\tau_R}\|}$. It is obvious that $\langle \bc_{\tau_{i+1},\dots,\tau_R}, \by_{\tau_{i+1},\dots,\tau_R}\rangle\le \langle \bc_{\tau_{i+1},\dots,\tau_R}, \lambda\bc_{\tau_{i+1},\dots,\tau_R}\rangle$ for all $\|\by_{\tau_{i+1},\dots,\tau_R}\|=\sqrt{r^2-\sum_{m=1}^ia^2_{\tau_m}}$. In other words, to maximize the inner product, $\by_{\tau_{i+1},\dots,\tau_R}$ should have the same direction as $\bc_{\tau_{i+1},\dots,\tau_R}$. Thus, we only need to show that this direction is feasible for $r\in\Gamma_i$. This is equivalent to show that $\lambda\bc_{\tau_{i+1},\dots,\tau_R}\in\Lambda_{\tau_{i+1},\dots,\tau_R}$ for any $r\in\Gamma_i$. We can easily prove that 
\[
r\in \Gamma_i \Leftrightarrow \lambda\in \Omega_i ,
\]
where $\Omega_i=\left[\phi_{\tau_i}^{-1},\phi_{\tau_{i+1}}^{-1}\right]$ for $i=0,\dots,R-1$. 
Thus, for any $r\in \Gamma_i$ and $j=\tau_{i+1},\dots,\tau_R$, we have $0\le \lambda c_j\le \phi_{\tau_{i+1}}^{-1} c_j\le \phi_j^{-1} c_j=a_{j}$. Hence, $\lambda\bc_{\tau_{i+1},\dots,\tau_R}\in\Lambda_{\tau_{i+1},\dots,\tau_R}$.
\end{proof}

Combining Lemma \ref{lemma-edge} and Lemma \ref{lemma-scale}, we have 
\begin{equation}
z^{(i)}_j=\left\{\begin{array}{l} a_j \hspace{10mm} j=\tau_1,\dots, \tau_i \\ %\frac{\sqrt{r^2-\sum_{j=1}^ia_{\tau_j}^2}}{\|\bc_{\tau_{i+1},\dots,\tau_R}\|} 
\lambda c_j \hspace{8mm} j=\tau_{i+1},\dots, \tau_R
\end{array}\right.
\label{zi}
\end{equation}
and thus
\begin{equation}
\max_{\|\by\|=r\in\Gamma_i,\by\in\Lambda}{\langle \bc, \by\rangle}=\sum_{m=1}^ib_{\tau_m}+\lambda \|\bc_{\tau_{i+1},\dots,\tau_R}\|^2.
\label{inner-solution}
\end{equation}

We have solved the inner optimization of Subproblem $i$. The solution of the $R$ subproblems can thus be obtained.

\begin{lemma}
For $i=1,\dots,R$, define
\[
\lambda_i=\frac{1+\sum_{m=1}^{i}a_{\tau_m}^2}{\sum_{m=1}^{i}b_{\tau_m}}.
\]
The solution of Subproblem $0$ is $\by^{(0)}=\phi_{\tau_1}^{-1}\bc$. The solution of Subproblem $i$ for $i=1,\dots,R-1$ is $\by^{(i)}$ that is defined as
\begin{equation}
y^{(i)}_j=\left\{\begin{array}{l} a_j \hspace{35mm} j=\tau_1,\dots, \tau_{i} \\ \min\left\{\lambda_i,\phi_{\tau_{i+1}}^{-1}\right\} c_{j} \hspace{8mm} j=\tau_{i+1},\dots,\tau_R.
\end{array}\right.
\label{sub-solution}
\end{equation}
\label{lemma-problem-i}
\end{lemma}

\begin{proof}
From (\ref{inner-solution}), Subproblem $i$ is equivalent to the following 1-dimensional optimization problem:
\begin{equation}
\max_{\lambda\in\Omega_i}\frac{\left(\sum_{m=1}^ib_{\tau_m}+ \|\bc_{\tau_{i+1},\dots,\tau_R}\|^2\lambda\right)^2}
{1+\sum_{m=1}^ia^2_{\tau_m}+\|\bc_{\tau_{i+1},\dots,\tau_R}\|^2\lambda^2}.
\label{decom-problem-1}
\end{equation}

When $i=0$, (\ref{decom-problem-1}) is equivalent to 
$\max_{\lambda\in\Omega_i}\frac{\|\bc\|^4\lambda^2}{1+\|\bc\|^2\lambda^2}$. Since $\frac{\|\bc\|^4\lambda^2}{1+\|\bc\|^2\lambda^2}$ is an increasing function of $\lambda$, its maximum is at $\lambda=\phi_{\tau_1}^{-1}$.

For $i=1,\dots,R-1$, Define 
\[
\xi_i(\lambda)=\frac{\left(\sum_{m=1}^ib_{\tau_m}+ \|\bc_{\tau_{i+1},\dots,\tau_R}\|^2\lambda\right)^2}
{1+\sum_{m=1}^ia^2_{\tau_m}+\|\bc_{\tau_{i+1},\dots,\tau_R}\|^2\lambda^2}.
\]
We have,
\[
\frac{\partial \xi_i}{\partial\lambda}=
\frac{2\left(\sum_{m=1}^ib_{\tau_m}+\|\bc_{\tau_{i+1},\dots,\tau_R}\|^2\lambda\right)\|\bc_{\tau_{i+1},\dots,\tau_R}\|^2}
{\left(1+\sum_{m=1}^ia^2_{\tau_m}+\|\bc_{\tau_{i+1},\dots,\tau_R}\|^2\lambda^2\right)^2}
\left(1+\sum_{m=1}^ia^2_{\tau_m}- \sum_{m=1}^ib_{\tau_m}\lambda\right).
\]
Thus, $\frac{\partial \xi_i}{\partial \lambda}>0$ if $\lambda<\lambda_i$ and $\frac{\partial \xi}{\partial \lambda}<0$ if $\lambda>\lambda_i$. So, if $\lambda_i\le \phi_{\tau_{i+1}}^{-1}$, the optimal solution is reached at $\lambda=\lambda_i$. Otherwise, the optimal solution is reached at  $\lambda=\phi_{\tau_{i+1}}^{-1}$. From (\ref{zi}), Subproblem $i$ is solved at $\by^{(i)}$ as defined in (\ref{sub-solution}).
\end{proof}

Now, we can work on the relay power control problem presented in (\ref{opt-problem}). 
\begin{theorem} Define $\bx^{(i)}$ as
\begin{equation}
x^{(i)}_j=\left\{\begin{array}{l} 1 \hspace{9mm} j=\tau_1,\dots, \tau_{i} \\ 
\lambda_i\phi_{j} \hspace{4mm} j=\tau_{i+1},\dots,\tau_R
\end{array}\right..
\label{sub-solution-x}
\end{equation}
The solution of the SNR optimization is $\bx^{(i_0)}$, where $i_0$ is the smallest $i$ such that $\lambda_i<\phi_{\tau_{i+1}}^{-1}$.
\label{thm-main}
\end{theorem}

\begin{proof} First, since $\phi_{R+1}=0$, we have $\lambda_R<\phi_{\tau_{R+1}}^{-1}=\phi_{R+1}^{-1}=\infty$. Thus, $i_0$ exists. Also, since 
$\lambda_{i_0}<\phi_{\tau_{i_0+1}}^{-1}$, and $\phi_{\tau_{j}}$ decreases with $j$, we have $x_{j}^{(i_0)}\le 1$ for $j=\tau_{i_0+1},\dots,\tau_{R}$. This means that $\bx^{(i_0)}$ is in the feasible region of the optimization problem.

Denote \[
\eta(\by)=\frac{\langle\bc,\by \rangle^2}{1+\|\by\|^2}.
\]
Note that $\|\by^{(0)}\|=r_1$. Since $r_1\in\Gamma_1$, $\by^{(0)}$ is also a feasible point of Subproblem 1. Thus, $\eta\left(\by^{(0)}\right)\le\eta\left(\by^{(1)}\right)$ due the optimality of $\by^{(1)}$ in Subproblem 1. This means that there is no need to consider Subproblem 0. 
%Therefore, to solve the optimization problem, we only need to check $\by^{(1)},\dots,\by^{(R-1)}$ and find the one with the largest $\eta\left( \by^{(i)}\right)$. However, 
For $i=1,\dots,R-2$, if $\lambda_i\ge \phi_{\tau_{i+1}}^{-1}$,
\[
y^{(i)}_j=\left\{\begin{array}{l} a_j \hspace{16mm} j=\tau_1,\dots, \tau_{i} \\ 
\phi_{\tau_{i+1}}^{-1}c_{j} \hspace{8mm} j=\tau_{i+1},\dots,\tau_R.
\end{array}\right.
\]
and 
\[
\|\by^{(i)}\|=\sqrt{\phi_{\tau_{i+1}}^{-2}\left\|\bc_{\tau_{i+1},\dots,\tau_R}\right\|^2+\sum_{j=1}^i |a_{\tau_j}|^2}=r_{i+1}.
\]
Since $r_{i+1}\in\Gamma_{i+1}$, $\by^{(i)}$ is a feasible point of Subproblem $i+1$. Thus, $\eta\left(\by^{(i)}\right)\le\eta\left(\by^{(i+1)}\right)$ due the optimality of $\by^{(i+1)}$ in Subproblem $i+1$. This means that there is no need to consider Subproblem $i+1$. Thus, we only need to check those $\by^{(i)}$'s with $\lambda_i<\phi_{\tau_{i+1}}^{-1}$, and find the one that results in the largest receive SNR. From the definition in (\ref{sub-solution-x}), this is the same as to check those $\bx^{(i)}$'s with $\lambda_i<\phi_{\tau_{i+1}}^{-1}$.

Now, we prove that $\lambda_{i+1} < \phi_{\tau_{i+2}}^{-1}$ if $\lambda_i < \phi_{\tau_{i+1}}^{-1}$. First, from $\lambda_i < \phi_{\tau_{i+1}}^{-1}$, we have
\[\frac{1+\sum_{m=1}^{i}a_{\tau_m}^2}{\sum_{m=1}^{i}b_{\tau_m}}< \phi_{\tau_{i+1}}^{-1}.\]
Since $\frac{a_{\tau_{i+1}}^2}{b_{\tau_{i+1}}}=\phi_{\tau_{i+1}}^{-1}$,
we can prove easily that
\[
\lambda_{i+1}=\frac{1+\sum_{m=1}^{i+1}a_{\tau_m}^2}{\sum_{m=1}^{i+1}b_{\tau_m}}
=\frac{1+\sum_{m=1}^{i}a_{\tau_m}^2+a_{\tau_{i+1}}^2}{\sum_{m=1}^{i}b_{\tau_m}+b_{\tau_{i+1}}}< \phi_{\tau_{i+1}}^{-1}<\phi_{\tau_{i+2}}^{-1}.\]
Thus, we only need to check those $\bx^{(i)}$'s for $i_0\le i \le R$ and find the one causing the largest receive SNR. From previous discussion, $i_0\ge 1$.

Define $SNR_i=\frac{\langle \bb, \bx^{(i)}\rangle^2}{1+\|A\bx^{(i)}\|^2}$. Now, we prove that $SNR_{i}>SNR_{i+1}$ for $i_0\le i \le R$. From the proof of Lemma 3, we have 
\setlength{\arraycolsep}{0pt}
\begin{eqnarray*}
SNR_i&=&\frac{\left(\sum_{m=1}^ib_{\tau_m}+ \|\bc_{\tau_{i+1},\dots,\tau_R}\|^2\lambda_i\right)^2}
{1+\sum_{m=1}^ia^2_{\tau_m}+\|\bc_{\tau_{i+1},\dots,\tau_R}\|^2\lambda_i^2} \\
%&=& \sum_{m=i+1}^R c_{\tau_{m}}^2+\left(1+\sum_{m=1}^ia_{\tau_m}^2\right) \lambda_i^{-2} \\
&=&\sum_{m=i+1}^R c_{\tau_{m}}^2+\frac{\left(\sum_{m=1}^{i}b_{\tau_m}\right)^2}{1+\sum_{m=1}^ia_{\tau_m}^2}  \\
&=&SNR_{i+1}+\frac{b_{\tau_{i+1}}^2}{a_{\tau_{i+1}}^2}+\frac{\left(\sum_{m=1}^{i}b_{\tau_m}\right)^2}{1+\sum_{m=1}^ia_{\tau_m}^2}-\frac{\left(\sum_{m=1}^{i+1}b_{\tau_m}\right)^2}{1+\sum_{m=1}^{i+1}a_{\tau_m}^2} \\
%&=&SNR_{k+1}+\frac{\left[b_{\tau_{k+1}}\left(1+\sum_{m=1}^ka_{\tau_m}^2\right)-\left(\sum_{m=1}^{k}b_{\tau_m}\right)a_{\tau_{k+1}}^2 \right]^2} {a_{\tau_{k+1}}^2\left(1+\sum_{m=1}^ka_{\tau_m}^2\right)\left(1+\sum_{m=1}^{k+1}a_{\tau_m}^2\right)}\\
&=&SNR_{i+1}+\frac{\left(1+\sum_{m=1}^{i}a_{\tau_m}^2\right) a^2_{\tau_{i+1}}} {1+\sum_{m=1}^{i+1}a_{\tau_m}^2}\left(\phi_{i+1}-\lambda_{i+1}^{-1} \right)^2\\
&>&SNR_{i+1}.
%&=&\sum_{m=k+2}^R c_{\tau_{m}}^2+\left(1+\sum_{m=1}^{k+1}a_{\tau_m}^2\right) \lambda_{k+1}^{-2}+c_{\tau_{k+1}}^2-a_{\tau_{k+1}}^2\lambda_{k+1}^{-2} +\left(1+\sum_{m=1}^ka_{\tau_m}^2\right)\left(\lambda_k^{-2}-\lambda_{k+1}^{-2}\right) \\
%&=&SNR_{k+1}+a_{\tau_{k+1}}^2\left(\phi_{\tau_{k+1}}^2-\lambda_{k+1}^{-2}\right) +\left(1+\sum_{m=1}^ka_{\tau_m}^2\right)\left(\lambda_k^{-2}-\lambda_{k+1}^{-2}\right)\\
%&>& SNR_{k+1}+\left(1+\sum_{m=1}^{k+1}a_{\tau_m}^2\right)\left(\lambda_k^{-2}-\lambda_{k+1}^{-2}\right) \hspace{3mm} 
%\mbox{(since $\lambda_k<\phi_{\tau_{k+1}}^{-1}$)}.
\end{eqnarray*}
\setlength{\arraycolsep}{5pt}
Thus, the optimal power control vector that maximizes the receive SNR is $\bx^{(i_0)}$.
%From (\ref{sub-solution}), (\ref{sub-solution-x}), (\ref{flag-func}), and $I_i=1$, we have $\bx^{(i)}=A^{-T}\by^{(i)}$. To maximize the receive SNR, we only need to check those $\bx^{(i)}$'s for $i=1,\dots,R-1$ with $I_i=1$ and find the one with the largest receive SNR. Thus, (\ref{opt-x}) is obtained.
\end{proof}

\subsection{Discussion}
\label{sec-PL-discussion}
It is natural to expect the power control at relays to undergo an on-or-off scenario: a relay uses its maximum power if its channels are good enough and otherwise not to cooperate at all.  Our result shows otherwise. The optimal power used at a relay can be any value between 0 and its maximal power. In many situations, a relay should use partial of its power, whose value is determined not only by its own channels but all others' as well. This is because every relay has two effects on the transmission. For one, it helps the transmission by forwarding the information, while for the other, it harms the transmission by forwarding noise as well. Its transmit power has a non-linear effect on the powers of both the signal and the noise, which makes the optimization solution not an on-or-off one, not a decoupled one, and, in general, not even a differentiable function of channel coefficients. 

As shown in Theorem \ref{thm-main} and Lemma \ref{lemma-problem-i}, the fraction of power used at relay $j$ satisfies $\alpha_{j}=1$ for $j=\tau_1,\dots,\tau_{i_0}$ and $\alpha_{j}=\lambda_{i_0}\phi_j$ for $j=\tau_{i_0+1},\dots,\tau_R$. Thus, the $i_0$ relays whose $\phi$'s are the largest use their maximal powers. Since $i_0\ge1$, there is at least one relay that uses its maximum power. This tells us that the relay with the largest $\phi$ always uses its maximal power. The remaining $R-i_0$ relays whose $\phi$'s are smaller only use parts of their powers. For $j=\tau_{i_0+1},\dots,\tau_R$, the power used at the $j$th relay is  $\alpha_j^2P_j=\lambda_{i_0}^2\phi_j^{2}P_j=\lambda_{i_0}^2|f_j/g_j|^2\left(1+|f_j|^2P_0\right)$, which is proportional to $|f_j/g_j|^2\left(1+|f_j|^2P_0\right)$ since $\lambda_{i_0}$ is a constant for each channel realization. Although $P_j$ does not appear explicitly in the formula, it affects the decision of whether the $j$th relay should use its maximal power. Actually, in determining whether a relay should use its maximal power, not only do the channel coefficients and power constraint at this relay account, but also all other channel coefficients and power constraints. The power constraint of the transmitter, $P_0$, plays a roll as well. 

Due to these special properties of the optimal power control solution, it can be implemented distributively with each relay knowing only its own channel information. In the following, we propose two distributed strategies. One is for networks with a small number of relays, and the other is more economical in networks with a large number of relays. 
%Here, by distributed strategy, we mean the strategy in which each relay needs its own channel information only.

The receiver, which knows all channels, can solve the power control problem. When the number of relays, $R$, is small, the receiver broadcasts the indexes of the relays that use their full powers and the coefficient $\lambda_{i_0}$. If relay $j$ hears its own index from the receiver, it will use its maximal power to transmit during the second step. Otherwise, it will use power $\lambda_{i_0}^2|f_j/g_j|^2\left(1+|f_j|^2P_0\right)$. The bits needed for the feedback is 
\[
i_0\log R+B_1<R\log R +B_1,
\]
where $i_0$ is the number of relays that use their maximal powers and $B_1$ is the number of bits needed in broadcasting the real number $\lambda_{i_0}$. Instead, the receiver can also broadcast two real numbers: $\lambda_{i_0}$ and a real number $d$ that satisfies $\phi_{\tau_{i_0}}>d>\phi_{\tau_{i_0+1}}$. Relay $j$ calculates its own $\phi_j$. If $\phi_j>d$, relay $j$ uses its maximal power. Otherwise, it uses power $\lambda_{i_0}^2|f_j/g_j|^2\left(1+|f_j|^2P_0\right)$. The number of bits needed for the feedback is $2B_1$. Thus, when $R$ is large, this strategy needs less bits of feedback compared to the first one.

Networks with an aggregate power constraint $P$ on relays were analyzed in \cite{lar}. In this case, with the same notation in Section \ref{sec-PL-result}, $P_j=P$ and $\sum_{j=1}^R\alpha_j^2\le1$. The optimal solution is 
\[
\alpha_j=\frac{\frac{|f_jg_j|\sqrt{1+|f_j|^2P_0}}{|f_j|^2P_0+|g_j|^2P+1}}
{\sqrt{\sum_{m=1}^R\frac{|f_mg_m|^2(1+|f_m|^2P_0)}{(|f_m|^2P_0+|g_m|^2P+1)^2}}}.
\]
$\alpha_j$ is a function of its own channels $f_j,g_j$ only and an extra coefficient $c=\sqrt{\sum_{m=1}^R\frac{|f_mg_m|^2(1+|f_m|^2P_0)}{(|f_m|^2P_0+|g_m|^2P+1)^2}}$, which is the same for all relays. Therefore, this power allocation can be done distributively with the extra knowledge of one single coefficient $c$, which can be broadcasted by the receiver. In our case, every relay has a separate power constraint. This is a more practical assumption in sensor networks since every sensor or wireless device has its own battery power limit. The power control solutions of the two cases are totally different. 

If relay selection is used and only one relay is allowed to cooperate, it can be proved easily that we should choose the relay with the highest
\[
h_j=h(f_j,g_j,P_j)=\frac{P_j|f_jg_j|^2}{1+|f_j|^2P_0+|g_j|^2P_j}.
\]
We call $h$ the relay selection function since a relay with a larger $h_j$ results in a higher receive SNR. 
While all relays are allowed to cooperate, the concepts of the best relay and relay selection function are not clear. Since the power control problem is a coupled one, it is hard to measure how much contribution a relay has. As discussed before, in network beamforming, a relay with a larger $\phi_j$  does not necessarily use a larger power or has more contribution. But we can conclude that if $\phi_k>\phi_l$, the fraction of power used at relay $k$, $\alpha_k$, is no less than the fraction of power used at relay $l$, $\alpha_l$. It is worth to mention that in network beamforming, relays with larger enough $\phi$'s use their maximal powers no matter what their maximal powers are. Actually, it is not hard to see that if at one time channels of all relays are {\em good}, every relay should use its maximum power.

\subsection{Simulation Results}
\label{sec-PL-simulation}
\begin{figure}
\centerline{\subfigure[2-relay network]{\includegraphics[width=3in]{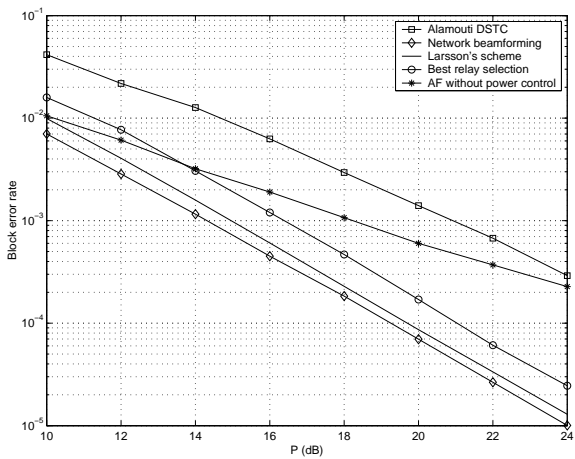}\label{fig-R2L2}}
\hfil
\subfigure[3-relay network]{\includegraphics[width=3in]{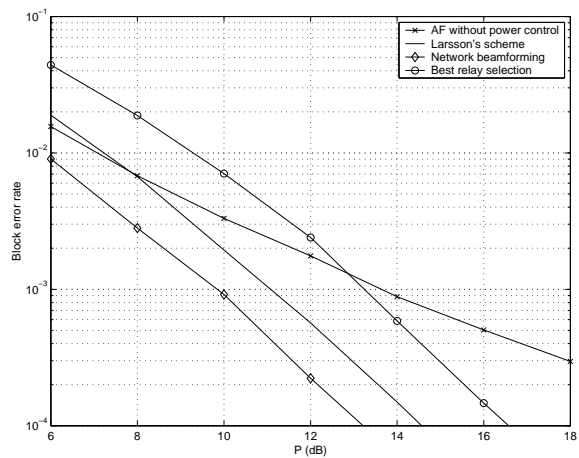}\label{fig-R3L2}}
}
\caption{Networks with fading channels and same power constraint for all nodes.}
%\label{fig-network-topology}
\end{figure}

In this section, we show simulated performance of network beamforming and compare it with performance of other existing schemes. Figures \ref{fig-R2L2} and \ref{fig-R3L2} show performance of networks with Rayleigh fading channels and the same power constraint on the transmitter and relays. In other words, $f_i,g_i$ are $\mathcal{CN}(0,1)$ and $P_0=P_1=\cdots=P_R=P$. The horizontal axis of the figures indicates $P$. In Fig.~\ref{fig-R2L2}, simulated block error rates of network beamforming with optimal power control are compared to those of best-relay selection, Larsson's scheme in \cite{lar} with total relay power $P$, distributed space-time coding in \cite{DSTC-paper}, and amplify-and-forward without power control (every relay uses its maximal power) in a 2-relay network.
%We assume that the power constraints at the transmitter and relays are all the same.  We assume Rayleigh fading channels, i.e., $f_i$ and $g_i$ are i.i.d.~$\mathcal{CN}(0,1)$. 
The information symbol $s$ is modulated as BPSK. We can see that network beamforming with optimal power control outperforms all other schemes. It is about 0.5dB and 2dB better than Larsson's scheme and best-relay selection, respectively. With perfect channel knowledge at relays, it is $7$dB better than Alamouti distributed space-time coding, which needs no channel information at relays. Amplify-and-forward with no power control only achieves diversity 1, distributed space-time coding achieves a diversity slightly less than two, while best-relay selection, network beamforming, and Larsson's scheme achieve diversity 2.  Fig.~\ref{fig-R3L2} shows simulated performance of a 3-relay network under different schemes. 
%Again, we assume that the power constraints at the transmitter and relays are all the same and are indicated by the horizontal axis of the figure.
Similar diversity results are obtained. But for the 3-relay case, network beamforming is about 1.5dB and 3.5dB better than Larsson's scheme and best-relay selection, respectively.

\begin{figure}
\centerline{\subfigure[Network with different relay powers ]{\includegraphics[width=3in]{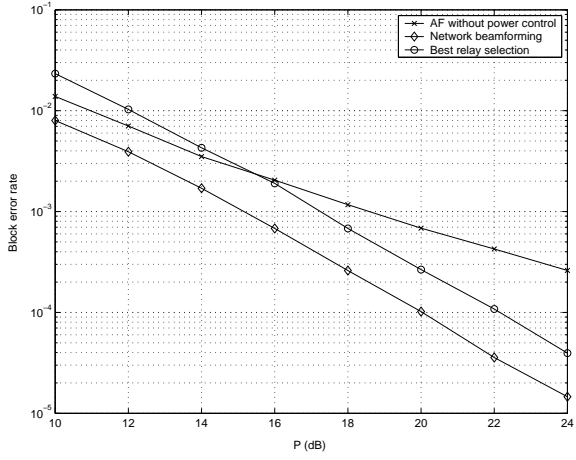}\label{fig-R2L2-DP}}
\hfil
\subfigure[Network with path-loss plus fading channels]{\includegraphics[width=3in]{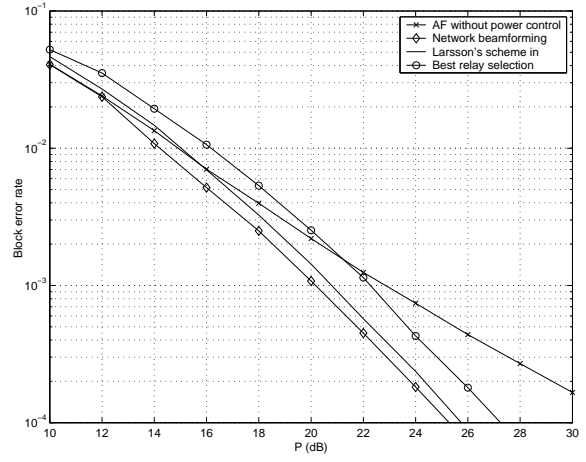}\label{fig-R2L2-Distance}}
}
\caption{2-relay networks with different relay powers and pass-loss plus fading channels.}
%\label{fig-network-topology}
\end{figure}

In Fig.~\ref{fig-R2L2-DP}, we show performance of a 2-relay network in which $P_0=P_1=P$ and $P_2=P/2$. That is, the transmitter and the first relay have the same power constraint while the second relay has only half the power of the first relay. The channels are assumed to be Rayleigh fading channels. In Fig.~\ref{fig-R2L2-Distance}, we show performance of a 2-relay network whose channels have both fading and path-loss effects. We assume that the distance between the first relay and the transmitter/receiver is 1, while the distance between the second relay and the transmitter/receiver is 2. The path-loss exponent \cite{Rappaport} is assume to be 2. We also assume that the transmitter and relays have the same power constraint, i.e., $P_0=P_1=P_2=P$. In both cases, distributed space-time coding does not apply, and Larsson's scheme applies for the second case only. So, we compare network beamforming with best-relay selection and amplify-and-forward with no power control only. Performance of Larsson's scheme is shown in Fig.~\ref{fig-R2L2-Distance} as well. Both figures show the superiority of network beamforming to other schemes.

\section{Networks with a Direct Link}
\label{sec-DL}
The previous section is on power control of relay networks with no DL between the transmitter and receiver. In this section, we discuss networks with a DL. As in \cite{NaBoKn}, there are several scenarios, which we discuss separately.

\subsection{Direct Link During the First Step Only}
\label{sec-DL-1ststep}
In this subsection, we consider relay networks with a DL during the first step only. This happens when the receiver knows that the transmitter is in vicinity and listens during the first step, while the transmitter is not aware of the DL or is unwilling to do the optimization because of its power and delay constraints. It can also happen when the transmitter is in the listening or sleeping mode during the second step.

In this case, $\beta_0=0$. From (\ref{x1}) and (\ref{x2}), the system equations can be written as
\setlength{\arraycolsep}{0.0em}
\[\left[\begin{array}{c} x_1 \\ x_2\end{array}\right]
= \left[\begin{array}{c} \alpha_0\sqrt{P_0}f_0 \\ \alpha_0\sqrt{P_0}\sum_{i=1}^R \frac{\alpha_i|f_ig_i|\sqrt{P_i}}{\sqrt{1+\alpha_0^2|f_i|^2P_0}} \end{array}\right]s
+\left[\begin{array}{c} w_1 \\ w_2+\sum_{i=1}^R \frac{\alpha_i|g_i|\sqrt{P_i}}{\sqrt{1+\alpha_0^2|f_i|^2P_0}}e^{-j\arg f_i}v_i \end{array}\right].
\]
\setlength{\arraycolsep}{0.5em}
Using maximum ratio combining, the ML decoding is
\setlength{\arraycolsep}{0.0em}
\begin{eqnarray*}
\arg\max_{s} \left|x_1- \alpha_0\sqrt{P_0}f_0s\right|^2+\left(1+\sum_{i=1}^R\frac{\alpha_i^2|g_i|^2P_i}{1+\alpha_0^2|f_i|^2P_0}\right)^{-1}
\left|x_2-\alpha_0\sqrt{P_0}\sum_{i=1}^R \frac{\alpha_i|f_ig_i|\sqrt{P_i}}{\sqrt{1+\alpha_0^2|f_i|^2P_0}}s\right|^2.
\end{eqnarray*}
\setlength{\arraycolsep}{0.0em}
The optimization problem is thus the maximization of the total receive SNR of both transmission branches, which equals
\[
\alpha_0^2P_0|f_0|^2+ \alpha_0^2P_0 \frac{\left(\sum_{i=1}^R \alpha_i\frac{|f_ig_i|\sqrt{P_i}}{\sqrt{1+\alpha_0^2|f_i|^2P_0}}\right)^2}{1+\sum_{i=1}^R\frac{\alpha_i^2|g_i|^2P_i}{1+\alpha_0^2|f_i|^2P_0}}.
\]
First, both terms in the SNR formula increase as $\alpha_0$ increases. Thus, $\alpha_0^*=1$, i.e., the transmitter should use its maximum power. The SNR optimization problem becomes the one in Section \ref{sec-PL-result}, in which there is no DL. Therefore, the power control of networks with a DL during the first step only is exactly the same as that of networks without a DL. This result is intuitive. Since with a DL during the first step only, operations at both the transmitter and relays keep the same as networks without the DL. The only difference is that the receiver obtains some extra information from the transmitter during the first step, and it can use the information to improve the performance without any extra cost. For the single-relay case, it can be proved easily that to maximize the receive SNR, the relay should use its maximal power as well, that is, $\alpha_1^*=1$.

\subsection{Direct Link During the Second Step Only}
\label{sec-DL-2ndstep}
In this subsection, we consider relay networks with a DL during the second step only. This happens when the transmitter knows that the receiver is at vicinity and determines to do more optimization to allocate its power between the two transmission steps. However, the receiver is unaware of the DL and is not listening during the first step. It can also happen when the receiver is in transmitting or sleeping mode during the first step.

In this case, $x_1=0$ and $x_2$ is given in (\ref{x2}). The receive SNR can be calculated to be
\[
\frac{P_0\left(\beta_0|f_0|+\alpha_0\sum_{i=1}^R\frac{\alpha_i|f_ig_i|\sqrt{P_i}}{\sqrt{1+\alpha_0^2|f_i|^2P_0}}\right)^2}
{1+\sum_{i=1}^R\frac{\alpha_i^2|g_i|^2P_i}{1+\alpha_0^2|f_i|^2P_0}}
\]
First, we show that $\alpha_0^2+\beta_0^2$ should take its maximal value 1, i.e., the transmitter should use all its power. Assume that $\hat{\alpha}_0^2+\hat{\beta}_0^2<1$ is the optimal solution. Define $\tilde{\beta}_0=\sqrt{1-\hat{\alpha}_0^2}$. We have $\tilde{\beta}_0>\hat{\beta}_0$. Therefore, $SNR(\hat{\alpha}_0,\hat{\beta}_0)<SNR(\hat{\alpha}_0,\tilde{\beta}_0)$. This contradicts the assumption that $(\hat{\alpha}_0,\hat{\beta}_0)$ is optimal.

Define 
\[
\hat{a}_i=\frac{|g_i|\sqrt{P_i}}{\sqrt{1+\alpha_0^2|f_i|^2P_0}}, \hspace{2mm} \hat{b}_i=\frac{\alpha_0|f_ig_i|\sqrt{P_i}}{\sqrt{1+\alpha_0^2|f_i|^2P_0}}, \hspace{2mm} 
\hat{c}_i=\frac{\hat{b}_i}{\hat{a}_i}, \hspace{2mm} 
\hat{A}=\diag\{\hat{\ba}\}, \hspace{2mm} \mbox{and} \hspace{2mm} \hat{y}_i=\hat{a}_i^{-1}\alpha_i.
\]
The receiver SNR can be calculated to be
\[
\psi(\alpha_0,\bx)=P_0\frac{\left(\sqrt{1-\alpha_0^2}|f_0|+\langle \hat{\bb}, \bx\rangle\right)^2}{1+\|\hat{A}\bx\|^2}
=P_0\frac{\left(\sqrt{1-\alpha_0^2}|f_0|+\langle \hat{\bc}, \hat{\by} \rangle\right)^2}{1+\|\hat{\by}\|^2}.
\]
For any fixed $\alpha_0$, we can optimize $\alpha_1,\dots,\alpha_R$ following the analysis in Section \ref{sec-PL-result}. The following theorem can be proved.

\begin{theorem} Define $\hat{\phi}_j=\frac{\hat{c}_{j}}{\hat{a}_{j}}$ for $j=1,\dots,R$ and $\hat{\phi}_{R+1}=0$. For any fixed $\alpha_0\in(0,1)$, order $\hat{\phi}_j$
%=\frac{|g_j|\sqrt{P_j}}{|f_j|\sqrt{1+\alpha_0^2|f_j|P_0}}$ 
as
\[
\hat{\phi}_{\hat{\tau}_1}\ge \cdots \ge \hat{\phi}_{\hat{\tau}_R}\ge \hat{\phi}_{\hat{\tau}_{R+1}}.
\]
For $i=0,\dots,R$, let $\hat{\lambda}_i=\frac{1+\sum_{m=1}^{i}\hat{a}_{\tau_m}^2}{\sqrt{1-\alpha_0^2}|f_0|+\sum_{m=1}^{i}\hat{b}_{\tau_m}}$ and define $\hat{\bx}^{(i)}$ is defined as
\[
\hat{x}^{(i)}_{j}=\left\{\begin{array}{l} 1 \hspace{13mm} j=\hat{\tau}_1,\dots,\hat{\tau}_i \\ 
\hat{\lambda}_i \hat{\phi}_j \hspace{7mm} j=\hat{\tau}_{i+1},\dots,\hat{\tau}_R \end{array}\right.
\]
The receive SNR is maximized at $\hat{\bx}^{(\hat{i}_0)}$, where $\hat{i}_0$ is the smallest $i$ such that $\hat{\lambda}_i<\hat{\phi}_{\hat{\tau}_{i+1}}^{-1}$.
\label{thm-main-DL-2nd}
\end{theorem}

\begin{proof} 
The proof of this theorem follows the one of Theorem \ref{thm-main} and the lemmas it uses.
\end{proof}

As discussed in Section \ref{sec-PL-result}, for networks with no DL, there is no need to consider the solution of Subproblem 0. Here it is different.
Define $\hat{r}_1=\hat{\phi}_{\hat{\tau}_1}^{-1}\|\hat{\bc}\|$. If we denote the solution of Subproblem 0, $\max_{|\hat{\by}|\in [0,\hat{r}_1],0_R\preceq \hat{\by}\preceq\hat{\ba}} \frac{\left(a+\langle \hat{\bc}, \hat{\by}\rangle\right)^2}{1+\|\hat{\by}\|^2}$, as $\hat{\by}^{(0)}$, because of the existence of the DL during the second step, it is possible that $\left\|\hat{\by}^{(0)}\right\|<\hat{r}_1$.
%Thus, we need to check one extra vector $\hat{\bx}^{(0)}=\hat{A}^{-T}\hat{\by}^{(0)}$ and binary function $\hat{I}_0$.

Now we discuss the optimization of $\alpha_0$. We first consider the case of $\alpha_0\in(0,1)$. For any given $\bx=\left[\begin{array}{ccc} \alpha_1 & \cdots & \alpha_R \end{array}\right]^T$, the $\alpha_0$ that maximizes 
the receive SNR satisfies $\frac{\partial \psi}{\partial \alpha_0}=0$. Thus, the optimal $\alpha_0$ can be found numerically by solving $\frac{\partial \psi}{\partial \alpha_0}=0$. It can be proved easily that $\frac{\partial \psi}{\partial \alpha_0}>0$ when $\alpha_0\rightarrow 0^+$ and $\frac{\partial \psi}{\partial \alpha_0}<0$ when $\alpha_0\rightarrow 1^-$. Thus, the maximum of $\psi$ is reached inside $(0,1)$.

When the power at the transmitter is high ($P_0\gg 1$), the receive SNR can be approximated by
\[
\psi(\alpha_0,\bx)\approx d(\alpha_0)=\frac{P_0\left(\sqrt{1-\alpha_0^2}|f_0|+d_1\right)^2}{1+d_2/\alpha_0^2},
\]
where $d_1=\frac{1}{\sqrt{P_0}}\sum_{i=1}^R\alpha_i|g_i|\sqrt{P_i}$ and $d_2=\frac{1}{P_0}\sum_{i=1}^R\alpha_i^2|g_i/f_i|^2P_i$. It can be calculated straightforwardly that for $\alpha_0\in(0,1)$,
\[
\frac{\partial d}{\partial \alpha_0} =\frac{4P_0\left(\sqrt{1-\alpha_0^2}|f_0|+d_1\right)}{\alpha_0^3\sqrt{1-\alpha_0^2}\left(1+d_2/\alpha_0^2\right)^2}
\left[-|f_0|\alpha_0^4-2b|f_0|\alpha_0^2+b|f_0|+ab\sqrt{1-\alpha_0^2}\right].
\]
and
%Thus, 
\[
\frac{\partial d}{\partial \alpha_0} =0 %&\Leftrightarrow & ab\sqrt{1-\alpha_0^2}=|f_0|\alpha_0^4+2b|f_0|\alpha_0^2-b|f_0| \\
\Leftrightarrow |f_0|^2\alpha_0^8-4d_2|f_0|^2\alpha_0^6+2d_2|f_0|^2\alpha_0^4+d_2^2(4|f_0|^2-d_1^2)\alpha_0^2+d_2^2(d_1^2-|f_0|^2)=0.
\]
This is a quartic equation of $\alpha_0^2$, whose solutions can be calculated analytically. Note that $\frac{\partial d}{\partial \alpha_0}>0$ when $\alpha_0\rightarrow 0^+$ and $\frac{\partial d}{\partial \alpha_0}<0$ when $\alpha_0\rightarrow 1^-$. Thus the maximum of $d$ is reached inside $(0,1)$. An approximate solution of $\alpha_0$ can thus be obtained analytically at high transmit powers.

Now we consider the cases of $\alpha_0=0$ and $\alpha_0=1$. If $\alpha_0=0$, the system degrades to a point-to-point one since only the DL works. Thus, the receive SNR is $|f_0|^2P_0$. For $\alpha_0=1$, we can obtain the optimal $\bx$ using Theorem \ref{thm-main-DL-2nd}. Thus, we obtain three sets of $\alpha$ and $\bx$ for the three cases: $\alpha_0\in(0,1)$, $\alpha_0=0$, and $\alpha_0=1$, respectively. The optimal solution of the system is the set of $\alpha_0$ and $\bx$ corresponding to the largest receive SNR. The power control problem in networks with a DL during the second step only can thus be solved using the following recursive algorithm.
\begin{algorithm} \hfil
\begin{enumerate}
\item Initialization: Set $\bx_1^{(previous)}=1_R$, the $R$-dimensional vector of all ones, $SNR_1^{(previous)}=0$, and $count=0$. Set the maximal number of iterations $iter$ and the threshold $thre$.
\item Optimize $\alpha_0$ with $\bx=\bx_1^{(previous)}$. Denote the solution as $\alpha_0^{(1)}$. We can either do this numerically or calculate the high SNR approximation. 
\item With $\alpha_0=\alpha_0^{(1)}$, find the $\bx$ that maximizes the receive SNR using Theorem \ref{thm-main-DL-2nd}. Denote it as $\bx_1$. Calculate $SNR_1=\psi(\alpha_0^{(1)},\bx_1)$.
%Calculate $SNR_1=\eta(\alpha_0^{(1)},\bx_1)$.
%\item Calculate the receiver SNR with the $\alpha_0$ obtained in Step 2 and the $\alpha_1,\dots,\alpha_R$ obtained in Step 3.
\item Set $count=count+1$. If $count<iter$ and $\left|SNR_1-SNR_1^{(previous)}\right|>thre$, set $\bx_1^{(previous)}=\bx_1$, $SNR_1^{(previous)}=SNR_1$, and go to step 2.
\item Find the solution of $\bx$ with $\alpha_0=1$ using Theorem \ref{thm-main-DL-2nd}. Denote this solution as $\bx_2$. 
\item The optimal solution is: $(\alpha_0^*,\bx^*)=\arg\max\left\{\psi(\alpha_0^{(1)},\bx_1),\psi(1,\bx_2),\psi(0,0_R)\right\}$.
\end{enumerate}
\label{alg-PL-DL-2nd}
\end{algorithm}

Similarly, the distributive strategies proposed in Section \ref{sec-PL-discussion} can be applied here.

\subsection{Direct Link During Both Steps}
\label{sec-DL-bothstep}
In this subsection, we consider relay networks with a DL during both the first and the second steps. This happens when both the transmitter and the receiver know that they are not too far away from each other and decide to communicate during both steps with the help of relays during the second step. 

From (\ref{x1}) and (\ref{x2}), the system equation can be written as
\setlength{\arraycolsep}{0.0em}
\[\left[\begin{array}{c} x_1 \\ x_2\end{array}\right]
= \left[\begin{array}{c} \alpha_0\sqrt{P_0}f_0 \\ \beta_0|f_0|+\alpha_0\sqrt{P_0}\sum_{i=1}^R \frac{\alpha_i|f_ig_i|\sqrt{P_i}}{\sqrt{1+\alpha_0^2|f_i|^2P_0}} \end{array}\right]s
+\left[\begin{array}{c} w_1 \\ w_2+\sum_{i=1}^R \frac{\alpha_i|g_i|\sqrt{P_i}}{\sqrt{1+\alpha_0^2|f_i|^2P_0}}e^{-j\arg f_i}v_i \end{array}\right].
\]
\setlength{\arraycolsep}{0.5em}
Similar to the networks discussed in Section \ref{sec-DL-1ststep}, the maximum ratio combining results in the following ML decoding:
%the noise covariance matrix is $\Pi$. With the same arguments, the ML decoding is
\begin{eqnarray*}
\arg\max_{s} \left|x_1- \alpha_0\sqrt{P_0}f_0s\right|^2+ \frac{\left|x_2-\sqrt{P_0}\left(\beta_0|f_0|+\alpha_0\sum_{i=1}^R \alpha_i\frac{|f_ig_i|\sqrt{P_i}}{\sqrt{1+\alpha_0^2|f_i|^2P_0}}\right)s\right|^2}
{\left(1+\sum_{i=1}^R\frac{\alpha_i^2|g_i|^2P_i}{1+\alpha_0^2|f_i|^2P_0}\right)^{-1}}.
\end{eqnarray*}
The total receive SNR of both transmission branches can be calculated to be 
\[
\alpha_0^2P_0|f_0|^2+ 
\frac{P_0\left(\beta_0|f_0|+\alpha_0\sum_{i=1}^R \frac{\alpha_i|f_ig_i|\sqrt{P_i}}{\sqrt{1+\alpha_0^2|f_i|^2P_0}}\right)^2}{1+\sum_{i=1}^R\frac{\alpha_i^2|g_i|^2P_i}{1+\alpha_0^2|f_i|^2P_0}} 
=\alpha_0^2P_0|f_0|^2+\psi(\alpha_0,\bx).
\]
The same as the networks in Section \ref{sec-DL-2ndstep}, $\alpha_0^2+\beta_0^2$ should take its maximal value, which is 1. That is, $\beta_0=\sqrt{1-\alpha_0^2}$. Similar to the SNR optimization in Section \ref{sec-DL-2ndstep}, for any given $\alpha_0\in(0,1)$, the SNR maximization  is the same as the maximization of $\psi$, which is solved by Theorem \ref{thm-main-DL-2nd}. But due to the difference in the receive SNR formula, the optimal $\alpha_0$ given $\alpha_1,\dots,\alpha_R$ is different. It is the solution of $2\alpha_0P_0|f_0|^2+\frac{\partial\psi}{\partial \alpha_0}=0$. When the DL exists during both steps, the case of $\alpha_0=0$, whose receive SNR is $|f_0|^2P_0$ will never outperform the case of $\alpha_0=1$, whose receive SNR is $|f_0|^2P_0+\psi(1,\bx)$ for some $\bx$. Thus, the case $\alpha_0=0$ needs not to be considered. The power control problem in networks with a DL during both steps can thus be solved using the following recursive algorithm.

\begin{algorithm} \hfil
\begin{enumerate}
\item Initialization: Set $\bx_1^{(previous)}=1_R$, $SNR_1^{(previous)}=0$, and $count=0$. Set the maximal number of iterations $iter$ and the threshold $thre$.
\item Optimize $\alpha_0$ with $\bx=\bx_1^{(previous)}$. Denote the solution as $\alpha_0^{(1)}$. We can do this numerically.
\item With $\alpha_0=\alpha_0^{(1)}$, find the $\bx$ that maximizes $\psi$ using Theorem \ref{thm-main-DL-2nd}. Denote it as $\bx_1$. 
Calculate $SNR_1=\left(\alpha_0^{(1)}\right)^2|f_0|^2P_0+\psi(\alpha_0^{(1)},\bx_1)$.
%\item Calculate the receiver SNR with the $\alpha_0$ obtained in Step 2 and the $\alpha_1,\dots,\alpha_R$ obtained in Step 3.
\item Set $count=count+1$. If $count<iter$ and $\left|SNR_1-SNR_1^{(previous)}\right|>thre$, set $\bx_1^{(previous)}=\bx_1$, $SNR_1^{(previous)}=SNR_1$ and go to step 2.
\item Find the solution of $\bx$ with $\alpha_0=1$ using Theorem \ref{thm-main-DL-2nd}. Denote this solution as $\bx_2$. 
\item The optimal solution is: $(\alpha_0^*,\bx^*)=\arg\max\left\{\left(\alpha_0^{(1)}\right)^2|f_0|^2P_0+\psi(\alpha_0^{(1)},\bx_1),|f_0|^2P_0+\psi(1,\bx_2)\right\}$.
\end{enumerate}
\label{alg-PL-DL-both}
\end{algorithm}

Again, the distributive strategies proposed in Section \ref{sec-PL-discussion} can be applied here.

\subsection{Performance Comparison}
In this subsection, we compare single-relay networks in which the power constraints at the transmitter and the relay are same, i.e., $P_0=P_1=P$. The channels are assumed to have both the fading and path-loss effect. There are four cases: no DL, a DL during the first step only, a DL during the second step only, and a DL during both steps.

\begin{figure}
\centerline{\subfigure[Triangle network]{\includegraphics[width=1.4in]{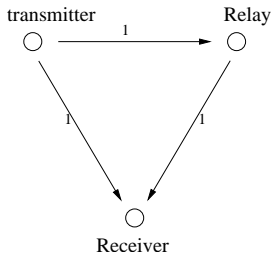}\label{fig-equallacteral-network}}
\hfil
\subfigure[Line network]{\includegraphics[width=2in]{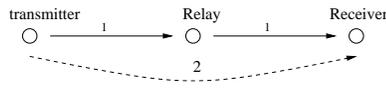}\label{fig-line-network}}
\hfil
\subfigure[Random network]{\includegraphics[width=2in]{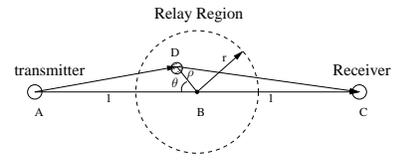}\label{fig-random-network}}
}
\caption{Network topology.}
%\label{fig-network-topology}
\end{figure}

\begin{figure}[t]
\centering
\includegraphics[width=4in]{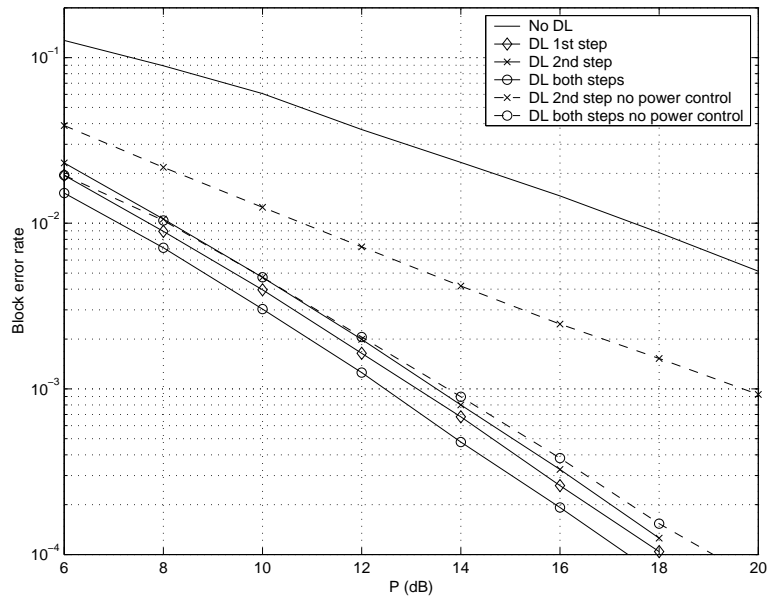}
\caption{Equilateral triangle network.}
\label{fig-triangle-network}
\end{figure}

In Fig.~\ref{fig-triangle-network}, we compare networks in which the distance of every link is the same, i.e., the three nodes are vertexes of an equilateral triangle with unit-length edges as shown in Fig.~\ref{fig-equallacteral-network}. We can see that the network with no DL has diversity 1 while networks with a DL and power control achieve diversity 2. The network with a DL during the first step performs less than 0.5dB better than the network with a DL during the second step only, while the network with a DL during both steps performs the best (about 1dB better than the network with a DL during the first step only). To illuminate the effect of power control, we show performance of networks whose transmit power at the relay and transmitter are fixed. For the network with a DL during the first step only, there is no power control problem since it is optimal for both the transmitter and the relay to use their maximal powers. For the other two cases, we let the transmitter uses half of its power, $P/2$, to each of the two steps and the relay always uses its maximum power $P$. We can see that, if the DL only exists during the second step, without power control, the achievable diversity is 1. At block error rates of $10^{-2}$ and $10^{-3}$, it performs 3 and 6dB worse, respectively. For networks with a DL during both steps, power control results in a 1.5dB improvement.

\begin{figure}[t]
\centerline{\subfigure[Path-loss exponent 2]{\includegraphics[width=3in]{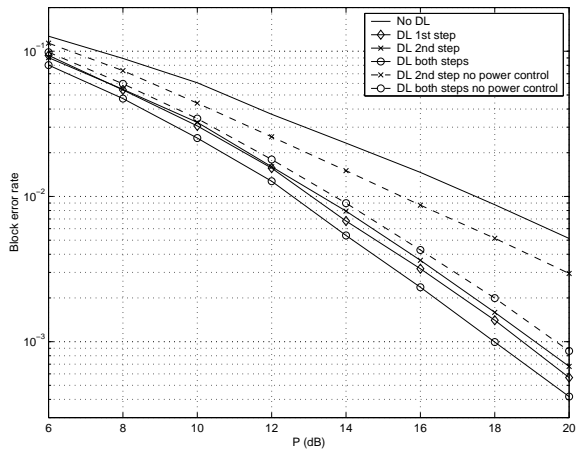}\label{fig-line-network-e2}}
\hfil
\subfigure[Path-loss exponent 3]{\includegraphics[width=3in]{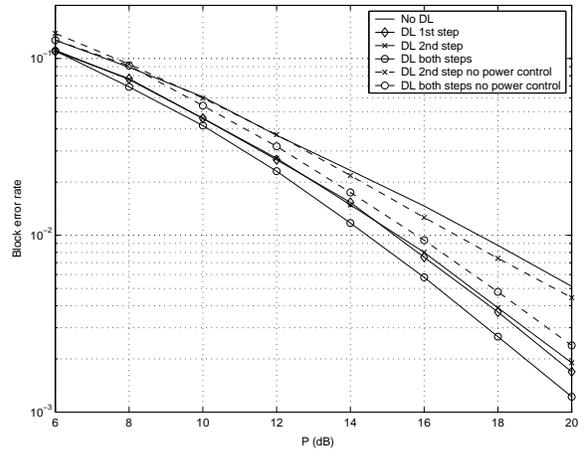}\label{fig-line-network-e3}}}
\caption{Single-relay line network.}
\end{figure}

In Fig.~\ref{fig-line-network-e2} and \ref{fig-line-network-e3}, we show performance of line networks with path-loss exponents 2 and 3 respectively. As shown in Fig.~\ref{fig-line-network}, the three nodes are on a line and the relay is in the middle of the transmitter and receiver.  The distance between the transmitter and receiver is assumed to be 2. The same phenomenon as in the equilateral triangle networks can be observed. The network with a DL during both steps performs the best (about 1dB better than the network with a DL during the first step only). The network with a DL at first step only performs slightly better than the one with a DL during the second step only. But the difference is smaller than that in Fig.~\ref{fig-triangle-network}. The performance difference between line networks with and without DLs is smaller than those in equilateral triangle networks, and it gets even smaller for larger path-loss exponents. This is because as the distance between the transmitter and receiver or the path-loss exponent is larger, the quality of the DL is lower. Therefore, the improvement due to this link is smaller. For both cases, power control results in a 1.5dB improvement when the DL link exists for both steps and a higher diversity when the DL exists for the second step only. 

Then we work on the random network in Fig.~\ref{fig-random-network}, in which the relay locates randomly and uniformly within a circle in the middle of the transmitter and the receiver. The distance between the transmitter and the receiver is assumed to be 2. The radius of the circle is denoted as $r$. We assume that $r< 1$. This is a reasonable model for ad hoc wireless networks since if communications between two nodes is allowed to be helped by one other relay, one should choose a relay that is around the middle of the two nodes. In other words, the distance between the relay and the transmitter or receiver should be shorter than that between the transmitter and receiver.
%The pass-loss exponent is assumed to be 2.

We work out the geometry first. As in Fig.~\ref{fig-random-network}, we denote the positions of the transmitter, the receiver, the relay, and the middle point of the transmitter and the receiver as $A, C, D$, and $B$, respectively. Denote the angle of $AB$ and $BD$ as $\theta$ and the length of $BD$ as $\rho$. The lengths of $AD$ and $CD$ are thus $\sqrt{1+\rho^2-2\rho\cos\theta}$ and $\sqrt{1+\rho^2+2\rho\cos\theta}$. Since $D$ is uniformly distributed within the circle, $\theta$ is uniform in $[0,\pi)$ and the pdf and cdf of $\rho$ can be calculated to be 
\[p(\rho)=\frac{2\rho}{r^2} \hspace{4mm} \mbox{and} \hspace{4mm} P(\rho<x)=\frac{x^2}{r^2},\]
%In simulations, we generate $\rho$ and $\theta$ according to the distributions. 
respectively. Define $Y=r\sqrt{X}$. If $X$ is uniform on $(0,1)$, it can be proved that 
\[
\Prob(Y\le x)=\Prob(r\sqrt{X}\le x)=\Prob\left(X\le \frac{x^2}{r^2}\right)=\frac{x^2}{r^2}.
\]
Thus, $Y$ has the same distribution as $\rho$. Therefore, we generate $Y$ to represent $\rho$.

\begin{figure}[t]
\centering
\includegraphics[width=4in]{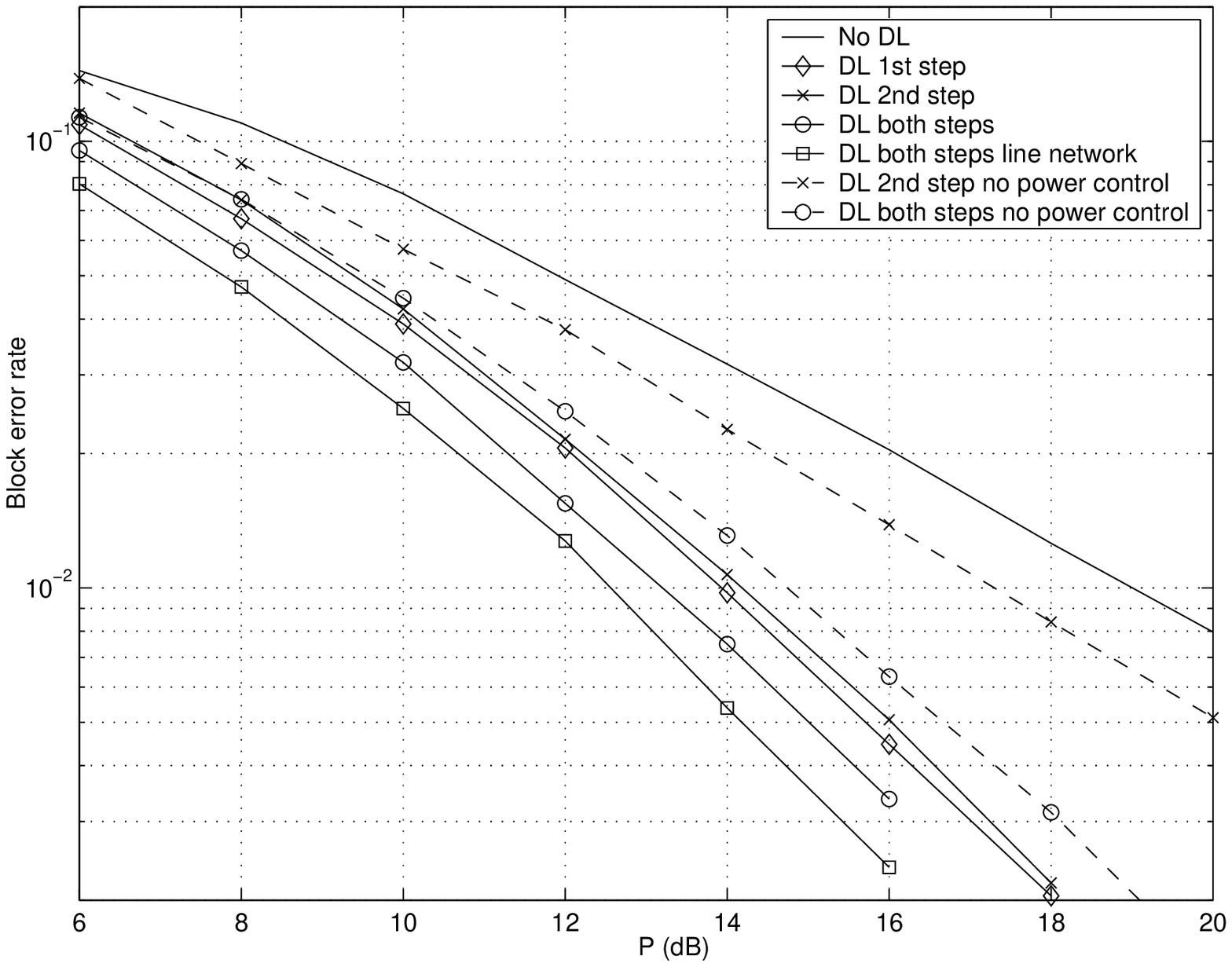}
\caption{Single-relay random network with pass-loss exponent 2 and $r=1/2$.}
\label{fig-random-network-P}
\end{figure}

Fig.~\ref{fig-random-network-P} shows performance of random networks with path-loss exponent 2 and $r=1/2$. We can see that the same phenomenon as in line networks can be observed. With a DL at both steps, the random network performs about 1dB worse than the line network. 

\section{Conclusions and Future Work}
\label{sec-conclusion}
In this paper, we propose the novel idea of beamforming in wireless relay networks to achieve both diversity and array gain. The scheme is based on a two-step amplify-and-forward protocol. We assume that each relay knows its own channels perfectly. Unlike previous works in network diversity, the scheme developed here uses not only the channels' phase information but also their magnitude. Match filters are applied at the transmitter and relays during the second step to cancel the channel phase effect and thus form a coherent beam at the receiver, in the mean while, optimal power control is performed based on the channel magnitude to decide the power used at the transmitter and relays. The power control problem for networks with any numbers of relays and no direct link is solved analytically. The solution can be obtained with a complexity that is linear in the number of relays. The power used at a relay depends on not only its own channels nonlinearly but also all other channels in the network. In general, it is not even a differentiable function of channel coefficients. Simulation with Rayleigh fading and path-loss channels show that network beamforming achieves the maximum diversity while amplify-and-forward without power control achieves diversity 1 only. Network beamforming also outperforms other cooperative strategies. For example, it is about 4dB better than best-relay selection. 

Relay networks with a direct link between the transmitter and receiver are also considered in this paper. For networks with a direct link during the first step only, the power control at relays and the transmitter is exactly the same as that of networks with no direct link. For networks with a direct link during the second step only and networks with a direct link during both steps, the solutions are different. Recursive numerical algorithms for the power control at both the transmitter and relays are given. Simulated performance of single-relay networks with different topologies shows that optimal power control results in about 1.5dB improvement in networks with a direct link at both steps and a higher diversity in networks with a direct link at the second step only.

We have just scratched the surface of a brand-new area. There are a lot of ways to extend and generalize this work. First,  it is assumed in this work that relays and sometimes the transmitter know their channels perfectly, which is not practical in many networks. Network beamforming with limited and delayed feedback from the receiver is an important issue. In multiple-antenna systems, beamforming with limited and delayed channel information feedback has been widely probed. However, beamforming in networks differs from beamforming in multiple-antenna systems in a couple of ways. In networks, it is difficult for relays to cooperate while in a multiple-antenna system, different antennas of the transmitter can cooperate fully. There are two transmission steps in relay networks while only one in multiple-antenna systems, which leads to different error rate and capacity calculation and thus different designs. Second, the relay network probed in this paper has only one pair of transmitter and receiver. When there are multiple transmitter-and-receiver pairs, an interesting problem is how relays should allocate their powers to aid different communication tasks. Finally, the two-hop protocol can be generalized as well. For a given network topology, one relevant question is how many hops should be taken to optimize the criterion at consideration, for example, error rate or capacity.

\newpage

{\small
\begin{thebibliography}{10}

\bibitem{Jbook}
H.~Jafarkhani, {\em Space-Time Coding: Theory and Practice}.
\newblock Cambridge Academic Press, 2005.

\bibitem{HTW}
A.~Hottinen, O.~Trikkonen, and R.~Wichman, {\em Multi-Antenna Transciever
  Techniques for 3G and Beyond}.
\newblock John Willey, 2003.

\bibitem{HuaMeiChangAsil}
Y.~Chang and Y.~Hua, ``Application of space-time linear block codes to parallel
  wireless relays in mobile ad hoc networks,'' in {\em Prof. of the 36th
  Asilomar Conference on Signals, Systems and Computers}, Nov. 2003.

\bibitem{HuaMeiChang}
Y.~Hua, Y.~Mei, and Y.~Chang, ``Wireless antennas - making wireless
  communications perform like wireline communications,'' in {\em Prof. of IEEE
  AP-S Topical Conference on Wireless Communication Technology}, Oct. 2003.

\bibitem{TangVal}
Y.~Tang and M.~C. Valenti, ``Coded transmit macrodiversity: block space-time
  codes over distributed antennas,'' in {\em Prof. of IEEE Vehicular Technology
  Conference 2001-Spring}, vol.~2, pp.~1435 -- 1438, May 2001.

\bibitem{SenErkipAa1}
A.~Sendonaris, E.~Erkip, and B.~Aazhang, ``User cooperation diversity-part {I}:
  System description,'' {\em IEEE Transactions on Communications}, vol.~51,
  pp.~1927--1938, Nov. 2003.

\bibitem{SenErkipAa2}
A.~Sendonaris, E.~Erkip, and B.~Aazhang, ``User cooperation diversity-part
  {II}: Implementation aspects and performance analysis,'' {\em IEEE
  Transactions on Communications}, vol.~51, pp.~1939--1948, Nov. 2003.

\bibitem{NaBoKn}
R.~U. Nabar, H.~B$\ddot{o}$lcskei, and F.~W. Kneubuhler, ``Fading relay channels:
  Performance limits and space-time signal design,'' {\em IEEE Journal on
  Selected Areas in Communications}, pp.~1099 -- 1109, Aug. 2004.

\bibitem{BoNaOyPa}
H.~B$\ddot{o}$lcskei, R.~U. Nabar, $\ddot{O}$.~Oyman, and A.~J. Paulraj, ``Capacity scaling laws
  in {MIMO} relay networks,'' {\em IEEE Transactions on Wireless
  Communications}, pp.~1433--1444, June 2006.

\bibitem{DSTC-paper}
Y.~Jing and B.~Hassibi, ``Distributed space-time coding in wireless relay
  networks,'' {\em IEEE Transactions on Wireless Communications}, Vol.~5, pp.~3524-3536, Dec.~2006.

\bibitem{DanaHassibi}
A.~F. Dana and B.~Hassibi, ``On the power-efficiency of sensory and ad-hoc
  wireless networks,'' {\em IEEE Transactions on Information Theory}, vol.~52, pp.~2890-2914, July 2006.

\bibitem{LenemanWornell}
J.~N. Laneman and G.~W. Wornell, ``Distributed space-time-coded protocols for
  exploiting cooperative diversity in wireless network,'' {\em IEEE Transactions on Information Theory}, vol.~49, pp.~2415-2425, Oct.~2003.

\bibitem{JHHN}
M.~Janani, A.~Hedayat, T.~E. Hunter, and A.~Nosratinia, ``Coded cooperation in
  wireless communications: space-time transmission and iterative decoding,''
  {\em IEEE Transactions on Signal Processing}, pp.~362-371, Feb. 2006.

\bibitem{DDSTC}
Y.~Jing and H.~Jafarkhani, ``Distributed differential space-time coding in
  wireless relay networks,'' {\em Accepted in IEEE Transactions on Communications}, 2006.

\bibitem{kira2}
T.~Kiran and B.~S. Rajan, ``Partial-coherent distributed space-time codes with
  differential encoder and decoder,'' in {\em Proc. of IEEE Internal Symposium
  on Information Theory}, 2006.

\bibitem{OgHa-allerton}
F.~Oggier and B.~Hassibi, ``A coding strategy for wireless networks with no
  channel information,'' in {\em Prof. of Allerton Conference}, 2006.

\bibitem{zz-eg-sc}
K.~Azarian, H.~E. Gamal, and P.~Schniter, ``On the achievable
  diversity-multiplexing tradeoff in half-duplex cooperative channels,'' {\em
  IEEE Transactions on Information Theory}, vol.~51, pp.~4152-4172, Dec.
  2005.

\bibitem{LaTsWo}
J.~N. Laneman, D.~N.~C. Tse, and G.~W. Wornell, ``Cooperative diversity in
  wireless networks: Efficient protocols and outage behavior,'' {\em IEEE
  Transactions on Information Theory}, pp.~3062-3080, Dec.~2004.

\bibitem{KaSh}
M.~Katz and S.~S. Shamai, ``Transmitting to colocated users in wireless ad hoc
  and sensory networks,'' {\em IEEE Transactions on Information Theory},
  pp.~3540-3563, Oct.~2005.

\bibitem{HuSaNo}
T.~E. Hunter, S.~Sanayei, and A.~Nosratinia, ``Outage analysis of coded
  cooperation,'' {\em IEEE Transactions on Information Theory}, pp.~375-391,
  Feb.~2006.

\bibitem{DeMiTa}
N.~Devroye, P.~Mitran, and V.~Tarokh, ``Achievable rates in cognitive radio
  channels,'' {\em IEEE Transactions on Information Theory}, pp.~1813-1827,
  May 2006.

\bibitem{LaCa}
E.~G. Larsson and Y.~Cao, ``Collaborative transmit diversity with adaptive
  resource and power allocation,'' {\em IEEE Communication Letters},
  pp.~511-513, June 2005.

\bibitem{CVS}
Y.~Cao, B.~Vojcic, and M.~Souryal, ``User-cooperative transmission with channel
  feedback in slow fading environment,'' in {\em Proc. of IEEE Vehicular
  Technology Conference 2004-Fall}, pp.~2063-2067, 2004.

\bibitem{lar}
P.~Larsson, ``Large-scale cooperative relaying network with optimal combining
  under aggregate relay power constraint,'' in {\em Proc. of Future
  Telecommunications Conference}, 2003.

\bibitem{AnKa}
P.~A. Anghel and M.~Kaveh, ``On the performance of distributed space-time
  coding systems with one and two non-regenerative relays,'' {\em IEEE
  Transactions on Wireless Communications}, pp.~682-692, Mar. 2006.

\bibitem{AKSA-Comm-06}
N.~Ahmed, M.~A. Khojastepour, A.~Sabharwal, and B.~Aazhang, ``Outage
  minimization with limited feedback for the fading relay channel,'' {\em IEEE
  Transactions on Communications}, pp.~659-699, Apr.~2006.

\bibitem{NLTW}
A.~Narula, M.~L. Lopez, M.~D. Trott, and G.~Wornell, ``Efficient use of side
  information in multiple-antenna data transmission over fading channels,''
  {\em IEEE Journals on Selected Areas in Communications}, pp.~1423-1436, Oct.~1998.

\bibitem{JaGo}
S.~A. Jafar and A.~J. Goldsmith, ``Transmit optimization and optimality of
  beamforming for multiple antenna systems with imperfect feedback,'' {\em IEEE
  Transactions on Wireless Communications}, pp.~1165-1175, July 2004.

\bibitem{msea}
K.~K. Mukkavilli, A.~Sabharwal, E.~Erkip, and B.~Aazhang, ``On beamforming with
  finite rate feedback in multiple-antenna systems,'' {\em IEEE Transactions on
  Information Theory}, pp.~2562- 2579, Oct.~2003.

\bibitem{rora}
J.~C. Roh and B.~D. Rao, ``Multiple antenna channels with partial channel state
  information at the transmitter,'' {\em IEEE Transactions on Wireless
  Communications}, pp.~677-688, Mar.~2004.

\bibitem{rora06}
J.~C. Roh and B.~D. Rao, ``Design and analysis of {MIMO} spatial multiplexing
  systems with quantized feedback,'' {\em IEEE Transactions on Signal
  Processing}, pp.~2874-2886, Aug.~2006.

\bibitem{ZhGi-mean}
S.~Zhou and G.~B. Giannakis, ``Optimal transmitter eigen-beamforming and
  space-time block coding based on channel mean feedback,'' {\em IEEE
  Transactions on Signal Processing}, pp.~2599-2613, Oct.~2002.

\bibitem{ZhGi-correlation}
S.~Zhou and G.~B. Giannakis, ``Optimal transmitter eigen-beamforming and
  space-time block coding based on channel correlations,'' {\em IEEE
  Transactions on Information Theory}, pp.~1673-1690, July 2003.

\bibitem{jso}
G.~Jongren, M.~Skoglund, and B.~Ottersten, ``Combining beamforming and
  orthogonal space-time block boding,'' {\em IEEE Transactions on Information
  Theory}, pp.~611-627, Mar.~2002.

\bibitem{Liu-Jar-SP}
L.~Liu and H.~Jafarkhani, ``Application of quasi-orthogonal space-time block
  codes in beamforming,'' {\em IEEE Transactions on Signal Processing},
  pp.~54-63, Jan.~2005.

\bibitem{LoHe}
D.~J. Love and R.~W. {Heath Jr.}, ``Limited feedback unitary precoding for orthogonal
  space-time block codes,'' {\em IEEE Transactions on Signal Processing},
  pp.~64-73, Jan.~2005.

\bibitem{EkJa}
S.~Ekbatani and H.~Jafarkhani, ``Combining beamforming and space-time coding
  for multi-antenna transmitters using noisy quantized channel direction
  feedback,'' {\em Submitted to IEEE Transactions on Communications}, 2006.

\bibitem{Rappaport}
T.~S. Pappaport, {\em Wireless Communications: Principles and Practice}.
\newblock Prentice Hall, 2nd~ed., 2002.

\end{thebibliography}
}
\end{document}